\documentclass[preprint,showpacs,preprintnumbers,amsmath,amssymb,floatfix]{revtex4-1}
\newcommand{\newwidth}{0.675\textwidth}
\newcommand{\newheight}{0.45\textwidth}

\usepackage{graphicx}
\usepackage{dcolumn,psfrag}
\usepackage{bm,verbatim}

\begin{document}

\title{
The formation of nonequilibrium steady states in interacting double quantum dots: 
When coherences dominate the charge distribution 
} 

\author{R.\ H\"artle$^{1,2}$}
\author{A.\ J.\ Millis$^{2}$}
\affiliation{
$^1$ Institut f\"ur theoretische Physik, Georg-August-Universit\"at G\"ottingen, Friedrich-Hund-Platz 1, D-37077 G\"ottingen, Germany. \\
$^2$ Department of Physics, Columbia University, New York, NY 10027, USA.  
}

\date{\today}

\begin{abstract}
We theoretically investigate the full time evolution of a nonequilibrium double quantum dot structure 
from initial conditions corresponding to different product states (no entanglement between dot and lead) to a 
nonequilibrium steady state. The structure is described by a two-level spinless Anderson model where the 
levels are coupled to two leads held at different chemical potentials. The problem is solved by a numerically 
exact hierarchical master equation technique and the results are compared to approximate ones obtained 
from Born-Markov theory. The methods allow us to study the time evolution up to times of order $10^4$ of the 
bare hybridization time, enabling eludication of the role of the initial state on the transient dynamics, 
coherent charge oscillations and an interaction-induced renormalization of energy levels. 
We find that when the system carries a single electron 
on average the formation of the steady state is strongly influenced by the coherence between the dots. 
The latter can be sizeable and indeed larger in the presence of a bias voltage than it is in equilibrium. 
Moreover, the interdot coherence is shown to lead to a pronounced difference in the population of the dots. 
\end{abstract}


\maketitle

\section{Introduction}

Understanding the time evolution of quantum mechanical systems is fundamentally important 
but in many cases also very challenging\cite{Grifoni1998,Haug98,May04,Fujisawa2006}. We probe 
quantum systems by following the time evolution induced by externally applied fields. Further 
the unique properties of driven quantum systems may be of technological 
importance \cite{Loss1998,Brumer03,Schoelkopf2008,Selinsky2013,Zwanenburg2013,Wu2014}. In many 
cases, the physics of interest involves interparticle interactions and possibly large departures 
from equilibrium. Therefore, it is essential to understand the nonlinear response of a quantum system in 
nonequilibrium situations. In these situations, however, analytical methods typically become too complex 
and numerical approaches are required. 

Quantum dot systems provide an important class of example systems to address nonlinear and 
nonequilibrium quantum physics\cite{Kastner2000,Aleiner2002,Reimann2002,Tews2004,Fujisawa2006,Andergassen2010}. 
Quantum dots are nanoscale regions in which electrons are spatially confined; they are often referred to as artificial atoms. 
The physics is thus characterized by a finite number of quantum mechanical degrees of freedom. However, unlike 
conventional atoms, quantum dots can easily be addressed by complex lead structures which provide both electron 
exchange (leading for example to transport through the dot) and the manipulation of each dot by electromagnetic 
fields \cite{Umansky1998,Holleitner2002,Wegscheider2007,WangPetta2013}. Moreover, their populations 
can reliably and non-invasively be read out using single-electron transistors or 
quantum point contacts \cite{Fujisawa2006,WangPetta2013,House2014}. 
As a result strongly nonequilibrium physics is 
accessible in the quantum dot context. Complex many-body phenomena such as, for example, 
Coulomb blockade\cite{Schmid1998,Kastner2000,Kiesslich2007,Brun2012} and Kondo 
correlations\cite{Cronenwett1998,Goldhaber1998,Liang02} are found even for the simplest quantum dot realization, 
namely a dot that can be characterized by a single spin-degenerate electronic level (even without spin-mixing effects 
such as, for example, in spin-valve setups \cite{Martinek2003,Braun2004,Rudzinski2005,Hell2014}). 
Quantum dots can be fabricated 
under well controlled conditions and in technologically scalable ways. The complexity and interest of the underlying 
physics increases with the number of levels on the dot and with the spatial structure enabled by larger dot structures. 
Therefore, they are suitable to study fundamental many-body phenomena \cite{Koenemann2006,Beckel2014} but can also be 
considered for electronic device applications such as, for example, solar energy conversion \cite{Selinsky2013,Wu2014} 
or quantum information processing \cite{Loss1998,Wiel2002,Sothmann2010,Baumgaertel2011,Zwanenburg2013}.

\begin{figure}
\begin{tabular}{cc}
a) \hspace{6.5cm} \text{ } & b) \hspace{8.5cm} \text{ }  \\[-0.2cm]
\includegraphics[width=0.5\textwidth]{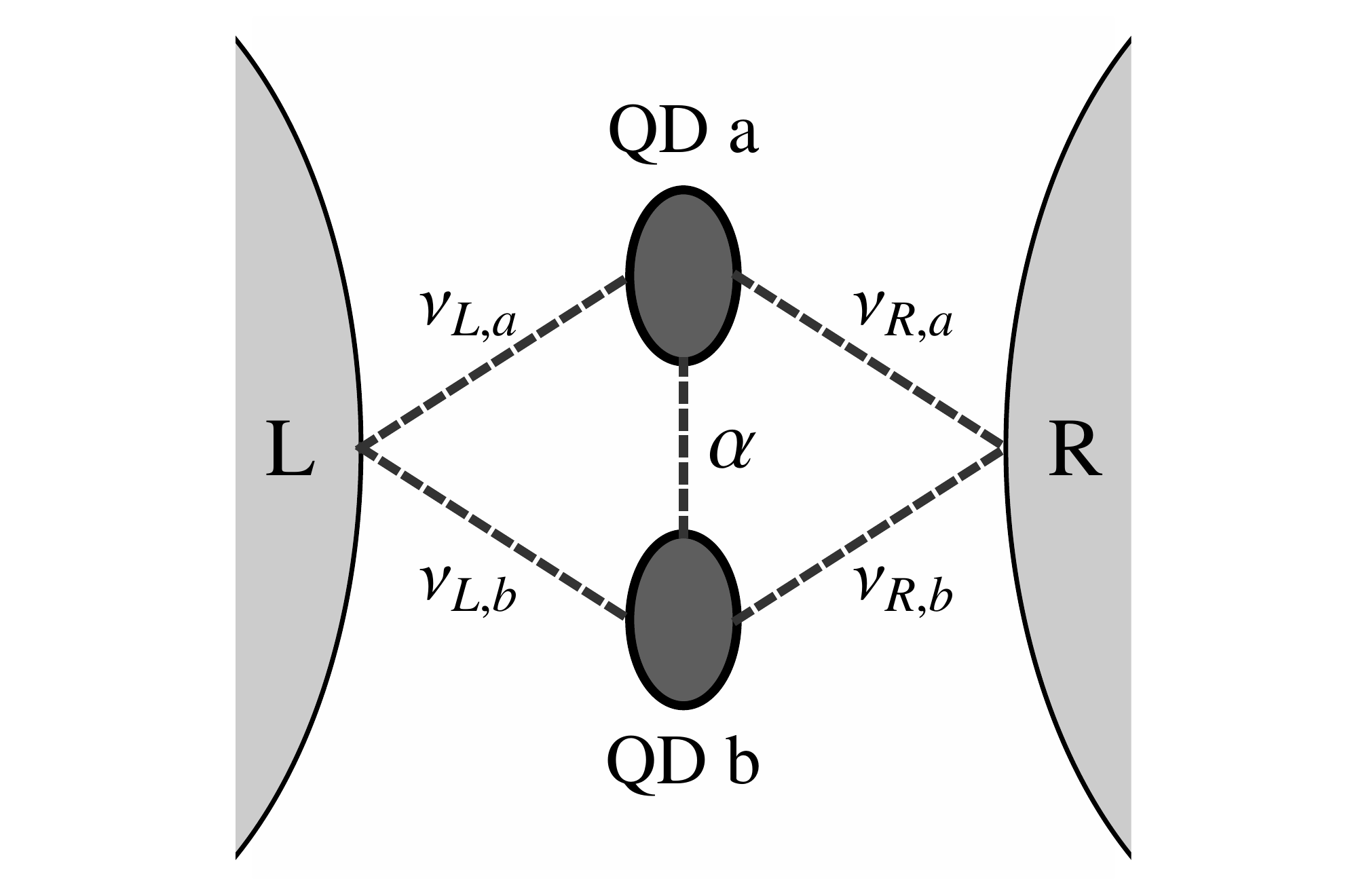} & \\[-11.4cm]
&\includegraphics[width=0.55\textwidth]{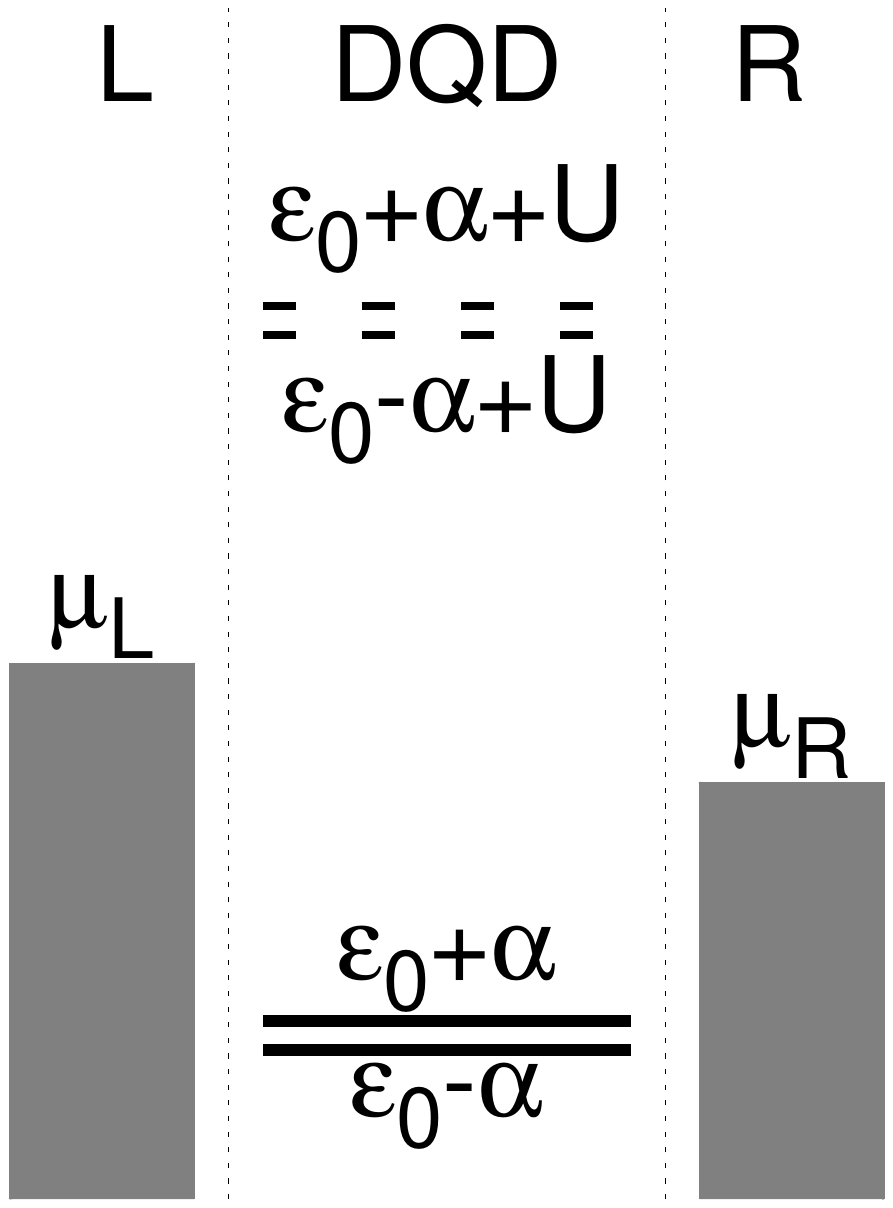} \\
\end{tabular}
\caption{\label{graphicsl} 
Panel a): Graphical representation of a double quantum dot system. The two quantum dots (QD) are coupled to a left (L) 
and a right electrode (R). The corresponding coupling matrix elements are denoted by $\nu_{K,m}$ with 
$K\in\{\text{L},\text{R}\}$ and $m\in\{a,b\}$. The inter-dot coupling is 
denoted by $\alpha$. In this work, we focus on a serial coupling configuration 
where $\nu_{\text{R},a}=\nu_{\text{L},b}=0$ and a branched configuration 
where $\nu_{\text{R},b}=\nu_{\text{L},b}=0$. 
Panel b): Level structure considered in this work 
where the single-particle levels are below and the levels associated with double occupancy are above the chemical 
potentials in the leads. This situation corresponds to a double quantum dot structure that is operated in 
the non-resonant transport regime. 
}
\end{figure}

In this work, we consider double quantum dot (DQD) structures \cite{Wiel2002} (cf.\ Fig.\ \ref{graphicsl}(a)). 
Mathematically, these structures may be thought of as two levels, coupled to each other and in a variety of possible 
ways to leads. They provide a simple model system for the examination of physics not accessible in transport through 
the widely-studied single-dot systems mentioned above,  
in particular sequential current flow from a lead into one dot, then into the other dot, and further 
into the other lead, but also internally gated situations where the occupancy of one dot affects flow through the other. 
In both of these cases the inter-dot coherence, which is defined as the off-diagonal element 
of the DQDs density matrix in the basis of the states localized on the quantum dots, 
will be seen to play a crucial role. 

In our analysis we suppress the spin degree of freedom, which is not essential to the physics of interest, and 
study the orbitally degenerate spinless Anderson model. This scenario can be experimentally realized by use of 
large magnetic fields or spin-polarized leads. The complexity of this problem is similar to that of a single 
quantum dot with a spin-degenerate level. Despite its simple structure, the orbitally degenerate spinless 
Anderson model exhibits a rich variety of complex many-body phenomena including 
orbital/pseudospin-Kondo physics \cite{Hofstatter2001,Kashcheyevs2007,Kubo2008,Trocha2012}, 
population inversion \cite{Lee2007,Goldstein2010,Karlstrom2011,Hartle2013b}, 
negative differential resistance \cite{Wunsch2005,Hettler2007,Trocha2009,RouraBas2010}, 
Fano-line shapes \cite{Goldstein2007,Wegscheider2007}, 
interaction-induced level repulsion\cite{Hofstatter2001,Wunsch2005} 
and resonances \cite{Meden2006,Lee2007,Schiller2009,Nilsson2010}. For some purposes one may think of the 
orbital degree of freedom in the spinless double-dot problem as playing a similar role as the 
spin degree of freedom in a conventional single-orbital dot\cite{Hofstatter2001,Kashcheyevs2007,Kubo2008,Trocha2012}. 
There is, however, a fundamental difference between the two systems: the coherence between the dots 
plays a key role in the double dot system, while it is typically zero or vanishes 
in the steady state of the aforementioned single-dot situations. 
In this paper we focus on phenomena that are associated with the dynamics of the coherence.

While substantial attention has been paid to the equilibrium and steady state properties of the two-orbital 
Anderson model, much less is known about the underlying dynamics. This deficiency in the literature can be traced back 
to the limitations of many time-dependent methods which involve approximations (giving, \emph{e.g.}, unphysical 
populations \cite{Egorova03,Kulkarni2013} and currents \cite{Hartle2013b}), impose severe restrictions on the 
accessible time scales 
\cite{Anderson05,Thorwart2008,Wang09,Werner2009,Segal2010,Schiro2010,Han2010,Muhlbacher2011,Gull2011,Weiss2013,Thoss2013,Cohen2013} 
or enable study only of parts of the full parameter space 
\cite{Anderson05,Jin2008,Thorwart2008,Segal2010,Timm2010,Segal2012,Weiss2013,Hartle2013b}.

In this article we analyse the nonequilibrium dynamics of the spinless double-dot system 
using the hierarchical quantum master equation (HQME) formalism \cite{Tanimura2006,Welack2006,Jin2008,Popescu2013,Hartle2013b}. 
In a previous paper \cite{Hartle2013b} we used the method to study the steady state transport properties of the 
spinless Anderson model, finding negative differential resistance due to interaction-induced decoherence \cite{Hettler2007} 
and explicating the role of interaction-induced energy-level renormalization \cite{Wunsch2005} in combination with level 
shifts related to the structure of the conduction bands \cite{Hartle2013b}. We found that these renormalization effects 
strongly affect the resonant transport properties of a quantum dot structure and give rise to lead-induced (RKKY-like) 
coupling effects \cite{Hartle2013b}.

Here, we analyze the time-dependence of the formation of a steady state from different (product) 
initial states under the influence of a time-independent Hamiltonian. 
As we will see, the transient dynamics strongly depends on the initial charge 
configuration in the dots, while the resulting steady state does not. 
Throughout this work, we focus on the non-resonant transport regime. As we have noted earlier in 
Ref.\ \onlinecite{Hartle2013b}, the associated time scales can be very long, because resonant tunneling processes 
are suppressed. The study is made possible by the HQME method, which facilitates a controlled (\emph{numerically exact}) 
study of this long-term dynamics. We will show that when the system carries a single electron on average, the corresponding 
transient dynamics exhibits a rich and complex behavior governed by a competition between exchange processes with the 
environment and coherent charge oscillations between the quantum dots. Both phenomena are strongly affected by an 
interaction-induced renormalization of the dots energy levels \cite{Wunsch2005,Darau2009,Donarini2010} that 
originates from exchange interactions with the electrodes. They are also known to give rise to a spin torque 
\cite{Martinek2003,Braun2004,Rudzinski2005} and a spin-precession resonance \cite{Hell2014} 
in spin-valve setups.

The HQME method \cite{Jin2008,Hartle2013b} allows us to obtain the time evolution 
of the double dot structure in a numerically exact way, assuming that the system is initially in a product state and 
that a systematic \cite{Hartle2013b} expansion in the hybridization of the system versus the temperature scale 
(which is set by the environment) converges. Internal consistency checks enable verification of the convergence. 
A significant advantage of the HQME method is the linear scaling of the numerical effort with the simulation time. 
This behavior is related to the time local formulation of the HQME  and makes it possible to reach simulation times 
greater than, \emph{e.g.}, a thousand times the inverse of the hybridization strength. This is essential in the present context 
and allows us to obtain reliable results for the effects of interest in this paper. 
Other numerically exact methods such as, for example, quantum Monte Carlo 
methods \cite{Werner2006,Thorwart2008,Muehlbacher08,Schmidt2008,Werner2009,Segal2010,Schiro2010,Gull2011,Weiss2013,Cohen2013}, 
time-dependent numerical renormalization group \cite{Anders2005,Anders2008,Korb2007} or density matrix renormalization 
group approaches  \cite{Schmitteckert04,Novotny2009,HeidrichMeisner2009} are not able to reach the needed timescales. 
Only reduced dynamics simulations \cite{Cohen2011}, either based on stochastic diagrammatic methods 
\cite{Cohen2013} or wave-function propagation schemes \cite{Wilner2013,Wilner2014}, can reach comparable time scales, 
provided that the corresponding memory kernel is decaying sufficiently fast.

In order to identify the physical mechanisms at work, we compare the exact results of the HQME scheme with approximate 
results that are obtained from Born-Markov theory \cite{May02,Mitra04,Lehmann04,Harbola2006,Volkovich2008,Hartle2010}. 
The standard Born-Markov approximation is related to HQME by (a) truncating the 
expansion at the lowest non-trivial order, (b) the Markov approximation and 
(c) the evaluation of the corresponding transition matrix elements 
(making a constant relaxation time approximation in the steady state). 
We therefore study two versions of the Born Markov approximation: 
the standard one and a modified version where we relax approximation (c). 
They are mainly distinguished by principal value terms, which encode 
the aforementioned renormalization effects. Thus, the effect of these terms 
can be visualized by comparing the two schemes. 
Although they enter the equation of motion of the coherence only, 
we find that the resulting coherent dynamics has also a strong influence on the population of the dots.

The article is organized in two parts. The first part (Sec.\ \ref{theory}) is devoted to the theoretical methodology. 
We briefly outline the model (Sec.\ \ref{modham}), the HQME approach (Sec.\ \ref{heomtheory}) and the two different 
Born-Markov schemes (Sec.\ \ref{bornmarkovtheory}). Results are presented in the second part of the 
article (Sec.\ \ref{results}). Throughout section \ref{results}, we focus on two complementary realizations of the 
spinless Anderson model: a serial and a branched configuration. A comparison of the two realizations will allow us 
to elucidate different aspects of the interaction-induced renormalization effects. Our analysis starts in 
Sec.\ \ref{nonmarkovian} with the time evolution from a nonequilibrium initial state to thermal equilibrium 
(\emph{i.e.}\ no bias voltage is applied to the quantum dots). In the subsequent section, Sec.\ \ref{feedback}, 
we compare these results to situations where a bias voltage is applied. We can therefore identify equilibrium and 
nonequilibrium effects in the formation of the steady state. Section \ref{conclusion} is a conclusion and the appendix 
includes technical details of the calculation.

\section{Theory}
\label{theory}

\subsection{Model Hamiltonian}
\label{modham}

We study the charge transfer dynamics of a biased double quantum dot (cf.\ Fig.\ \ref{graphicsl}(a)). We assume 
that each dot contains one electronic state and neglect spin degeneracy. Such a system can be realized by an array of 
quantum dots arranged to form an Aharonov-Bohm interferometer \cite{Holleitner2001,Iye2001,Holleitner2002,Wegscheider2007} 
or a nanoscale molecular conductor with an appropriate level structure 
\cite{Kalyanaraman2002,Ernzerhof2005,Solomon2008b,Brisker2010,Markussen2010,Lambert2011,Brisker2012,Hartle2011,Ballmann2012,Hartle2012}. 
The spinless situation may be realized physically if the spin degeneracy is lifted by an external magnetic field or by spin-polarized 
electrodes. The situation is modeled by a two-state spinless Anderson model 
\begin{eqnarray}
 H_{\text{DQD}} &=& \sum_{m\in\{a,b\}} \epsilon_{m} d_{m}^{\dagger}d_{m} + 
\alpha d_{a}^{\dagger}d_{b} +  \alpha d_{b}^{\dagger}d_{a} 
+ U d_{a}^{\dagger}d_{a} d_{b}^{\dagger}d_{b}. 
\end{eqnarray}
The dots are labelled by $a$ and $b$. The dot states are addressed by annihilation and creation operators $d_{a/b}$ 
and $d_{a/b}^{\dagger}$ with corresponding energies $\epsilon_{a/b}$. The inter-dot coupling strength is denoted by $\alpha$. 
A simultaneous population of the dots requires an additional charging energy $U>0$, reflecting repulsive Coulomb 
interactions between the electrons in the system. Note that this system is equivalent to a Kondo impurity 
if $\alpha\rightarrow0$.

The dynamics of the system is driven by charge exchange processes with the leads. The leads provide a reservoir of electrons, 
which can be described by a continuum of non-interacting electronic states 
\begin{eqnarray}
\label{hLR}
H_{\text{L/R}} &=& \sum_{k\in\text{L/R}} \epsilon_{k} c_{k}^{\dagger}c_{k}   
\end{eqnarray} 
with energies $\epsilon_{k}$ and corresponding annihilation and creation operators $c_{k}$ and $c_{k}^{\dagger}$. 
These continuum states are coupled to the states of the double dot system. The respective coupling operators can be written as 
\begin{eqnarray}
\label{htun}
H_{\text{tun}} &=& \sum_{k\in\text{L,R};m\in\{a,b\}} ( V_{mk} c_{k}^{\dagger}d_{m} + \text{h.c.} ). 
\end{eqnarray} 
The tunneling efficiency between the dots and the electrodes is given by the coupling matrix elements $V_{mk}$. 
It depends on the energy of the tunneling electrons and can be characterized by the level-width functions  
\begin{eqnarray}
\label{tuneff}
\Gamma_{K,mn}(\epsilon) &=& 2\pi\sum_{k\in K} V_{mk}^{*} V_{nk}\delta(\epsilon-\epsilon_{k})  
\end{eqnarray}
with $K\in\{\text{L,R}\}$. 

While the HQME formalism we discuss below applies for general dot-lead coupling, we will present results for two cases: 
the SERIAL configuration in which $V_{bk\in L}=V_{ak\in R}=0$ so that (for positive bias) 
current flows from the left lead into dot $a$, then from dot $a$ to dot $b$, and further from dot $b$ 
into the right lead, and the BRANCHED configuration in which $V_{bk}=0$ so that current flows through dot $a$ and 
dot $b$ is coupled to the leads only via its coupling to dot $a$.

If both electrodes have the same temperature $T$ and chemical potential $\mu$ the system will relax to a thermal 
equilibrium state. Departures from equilibrium may be induced by imposing a difference of temperature or chemical potential 
between the leads. We will typically assume that the lead temperatures are the same and induce nonequilibrium physics 
via a non-zero bias voltage, \emph{i.e.}\ $\Phi=\mu_{\text{L}}-\mu_{\text{R}}\neq0$. Throughout this work, we assume a 
symmetric drop of the bias voltage at the contacts, that is the chemical potentials of the left and the right leads are 
given by $\mu_{\text{L}}=-\mu_{\text{R}}=\Phi/2$. Note that this assumption is not decisive for our discussion.

The Hamiltonian of the whole system is given by 
\begin{eqnarray}
H &=& H_{\text{DQD}} + H_{\text{L}} + H_{\text{R}} + H_{\text{tun}}.     
\end{eqnarray}

\subsection{Hierarchical master equation approach} 
\label{heomtheory}

In order to determine the nonequilibrium dynamics of the double dot system, 
we employ the hierarchical quantum master equation method \cite{Tanimura2006,Welack2006,Jin2008,Popescu2013,Hartle2013b}. 
This is an equation of motion technique to determine the reduced density matrix 
\begin{eqnarray}
 \sigma(t) = \text{Tr}_{\text{L+R}}\left\{ \varrho(t) \right\},  
\end{eqnarray}
where the density matrix of the full system (\emph{i.e.}\ L--DQD--R) is denoted by $\varrho(t)$. 
A detailed derivation is given in Refs.~\onlinecite{Jin2008,Hartle2013b}. 
Here for completeness and to establish notation we review the derivation.

The equation of motion of the reduced density matrix\footnote{Note that the formalism is written in an interaction picture 
with respect to the lead Hamiltonians $H_{\text{L/R}}$. The dot Hamiltonian $H_{\text{DQD}}$ is explicitly excluded. 
This treatment allows us to suppress the direct appearance of dynamical phases in the reduced density matrix.} 
\begin{eqnarray}
 \frac{\text{d}}{\text{d}t} \sigma(t) &=& 
  -i \left[ H_{\text{DQD}} , \sigma(t) \right] 
 - \sum_{m,s} \left[ d_{m}^{s}, \tilde{\sigma}_{ms}(t) \right] 
\end{eqnarray}
is written in terms of a set of auxiliary operators 
\begin{eqnarray}
 \sum_{m,s} \left[ d_{m}^{s}, \tilde{\sigma}_{ms}(t) \right] &=& 
 i \text{Tr}_{\text{L+R}}\left\{ \left[ H_{\text{tun}}(t) , \varrho(t) \right]  \right\}  
\end{eqnarray}
with $s\in\{+,-\}$, $d_{n}^{+}=d_{n}^{\dagger}$ and $d_{n}^{-}=d_{n}$ and 
\begin{eqnarray}
 H_{\text{tun}}(t) &=& \text{e}^{i \left(H_{\text{L}} + H_{\text{R}} \right) t} H_{\text{tun}} 
 \text{e}^{-i \left(H_{\text{L}} + H_{\text{R}} \right) t}. 
\end{eqnarray} 
These operators encode the dynamics of the system that is induced by the coupling to the electrodes. 
They can be determined by a set of equations of motion. These equations lead, a priori, to another set 
of auxiliary operators, which are associated with the commutators 
$\left[ H_{\text{tun}}(t) , \left[ H_{\text{tun}}(t) , \varrho(t) \right] \right]$ and 
$\left[ \dot{H}_{\text{tun}}(t) , \varrho(t) \right]$. This can be continued, leading to a hierarchy of 
operators where the appearance of nested commutators such as 
$\left[ H_{\text{tun}}(t) , \left[ H_{\text{tun}}(t) , ..., \varrho(t) \right] \right]$ suggests the 
existence of a systematic expansion in terms of the hybridization operator $H_{\text{tun}}$. At this point, 
however, a hybridization expansion cannot be performed because of the operators that are associated with the 
time derivatives of the dot-lead coupling operator $\partial_{t}H_{\text{tun}}(t)$, $\partial^{2}_{t}H_{\text{tun}}(t)$, ...

A systematic approach to this problem is given in Refs.\ \cite{Jin2008,Hartle2013b}. It employs the correlation functions 
\begin{eqnarray}
C^{s}_{K,mn}(t-t') &=& 
\sum_{k\in K} V^{\overline{s}}_{mk} V_{nk}^{s} \text{Tr}_{K}\left\{ \sigma_{K} c_{k}^{s}(t) c_{k}^{\overline{s}}(t') \right\},  
\end{eqnarray}
where 
\begin{eqnarray}
 \sigma_{K} &=& 
 \frac{1}{\text{Tr}_{K}\left\{ \text{e}^{-\sum_{k\in K}\frac{\epsilon_{k}-\mu_{\text{L/R}}}{k_{\text{B}}T} c_{k}^{\dagger}c_{k}  }\right\} } 
 \text{e}^{-\sum_{k\in K}\frac{\epsilon_{k}-\mu_{\text{L/R}}}{k_{\text{B}}T} c_{k}^{\dagger}c_{k}}, 
\end{eqnarray}
$k_{\text{B}}$ denotes the Boltzmann constant, 
$\overline{s}=-s$, $V_{mk}^{+}=V_{mk}$, $V_{mk}^{-}=V_{mk}^{*}$, $c_{k}^{+}=c_{k}^{\dagger}$ and $c_{k}^{-}=c_{k}$. 
These functions characterize the tunneling processes between the dots and the electrodes.  
They are given by the tunneling efficiencies $\Gamma_{K,mn}(\omega)$ and the population of the electronic 
states in the leads, that is the respective Fermi distribution functions $f_{K}(\omega)$:  
\begin{eqnarray} 
 C^{s}_{K,mn}(t) &=& \int_{-\infty}^{\infty}\frac{\text{d}\omega}{2\pi}\, \text{e}^{si\omega t} 
 \Gamma_{K,mn}^{s}(\omega) f^{s}_{K}(\omega), 
\end{eqnarray}
with the short-hand notations 
$\Gamma_{K,mn}^{+}(\omega)=\Gamma_{K,mn}(\omega)$, $\Gamma_{K,mn}^{-}(\omega)=\Gamma_{K,nm}(\omega)$, 
$f_{K}^{+}(\omega)=f_{K}(\omega)$ and $f_{K}^{-}(\omega)=1-f_{K}(\omega)$. The auxiliary 
operators $\tilde{\sigma}_{ms}(t)$ can be written in terms of these correlation functions as \cite{Jin2008}
\begin{eqnarray}
\label{proboop}
 \tilde{\sigma}_{ms}(t) &=& 
 \sum_{Kn} \int_{0}^{t}\text{d}\tau\, C^{\overline{s}}_{K,mn}(t-\tau) 
 \text{Tr}_{\text{L+R}}\left\{ U(t,\tau) d_{n}^{\overline{s}} U(\tau,0) \varrho(0) U^{\dagger}(t,0)\right\}   \\
 && 
 - \sum_{Kn} \int_{0}^{t}\text{d}\tau\, C^{s,*}_{K,mn}(t-\tau) \text{Tr}_{\text{L+R}}\left\{ U(t,0) \varrho(0) 
U^{\dagger}(\tau,0) 
 d_{n}^{\overline{s}} U^{\dagger}(t,\tau) \right\},    \nonumber
\end{eqnarray} 
with the time evolution operator 
\begin{eqnarray}
 U(t,0) &=& T\left(\text{e}^{-i \int_{0}^{t}\text{d}\tau \left( H_{\text{tun}}(\tau) + H_{\text{DQD}} \right)}\right) .  
\end{eqnarray}
The formalism requires the assumption that the system is initially in a factorized state, 
\emph{i.e.}\ $\varrho(0)=\sigma(0)\sigma_{\text{L}}\sigma_{\text{R}}$.  The problem with the time derivatives of the 
dot-lead coupling operator is thus transferred to a representation of the time derivatives of the correlation 
functions $C^{\overline{s}}_{K,mn}$. The equations can be solved if we find a set of functions, which can be used 
to represent both the correlation functions  $C^{\overline{s}}_{K,mn}$ and its time derivatives.

Such a set of functions can be obtained, for example, by the Meir-Tannor parametrization 
scheme \cite{Tannor1999,Welack2006,Jin2008} for the tunneling efficiencies $\Gamma_{K,mn}(\epsilon)$ 
and the Pade approximation scheme for the Fermi distribution functions $f_{K}(\omega)$ \cite{Ozaki2007,Hu2010,Hu2011}. 
These \emph{sum-over-poles} schemes allow us to write the correlation functions $C^{s}_{K,mn}$ by a set of 
exponential functions \footnote{In general, in particular at zero temperature and large enough time scales, 
the correlation functions $C^{s}_{K,mn}(t)$ scale with the inverse of time, $\sim1/t$. Thus, 
a parametrization of $C^{s}_{K,mn}(t)$ in terms of exponentials works best at high temperatures.}
\begin{eqnarray}
\label{correxpand}
 C^{s}_{K,mn}(t) &=& \sum_{p} \eta^{s}_{K,mn,p} \text{e}^{-\omega^{s}_{K,p}t}, 
\end{eqnarray}
where the scheme to obtain the frequencies $\omega^{s}_{K,p}$ and the amplitudes $\eta^{s}_{K,mn,p}$ 
is outlined in the appendix. Corresponding to each of the exponential functions $\text{e}^{-\omega^{s}_{K,p}t}$, 
a new set of auxiliary operators can be defined as  
\begin{eqnarray}
\label{defauxop2} 
 \sigma_{K,mn,s,p}(t) &=& \eta^{s}_{K,mn,p} \int_{0}^{t}\text{d}\tau\,  \text{e}^{- \omega^{s}_{K,p} (t-\tau)} 
 \text{Tr}_{\text{L+R}}\left\{ U(t,\tau) d_{n}^{s} U(\tau,0) \varrho(0) U^{\dagger}(t,0)\right\}  \\
 && 
 - \eta^{\overline{s},*}_{K,mn,p} \int_{0}^{t}\text{d}\tau\, \text{e}^{- \omega^{s}_{K,p} (t-\tau)} 
 \text{Tr}_{\text{L+R}}\left\{  U(t,0) \varrho(0) U^{\dagger}(\tau,0) d_{n}^{s}
 U^{\dagger}(t,\tau) \right\}.  \nonumber
\end{eqnarray}
The time derivative of these operators involves only the operator itself (times the frequency $\omega^{s}_{K,p}$) 
and operators that contain an additional dot-lead coupling term $H_{\text{tun}}$. This allows us to establish a 
closed set of equations of motions in the sense that time derivatives do not lead to new classes of operators 
that are of the same order in $H_{\text{tun}}$. The operators $\sigma_{K,mn,s,p}(t)$ and the corresponding 
higher-tier operators can be written as 
\begin{eqnarray}
\label{defauxop} 
 \sigma^{(\kappa)}_{j_{1}..j_{\kappa}}(t) &=& \text{Tr}_{\text{L+R}}\left\{ B_{j_{\kappa}} .. B_{j_{1}} \varrho(t) \right\}, 
\end{eqnarray}
introducing superoperators $B_{j}$, 
\begin{eqnarray}
\text{Tr}_{\text{L+R}}\left\{ B_{j} \varrho(t) \right\} &\equiv& \sigma_{K,mn,s,p}(t), 
\end{eqnarray}
and superindices $j=(K,mn,s,p)$. By construction, the corresponding equations of motion 
\begin{eqnarray}
\label{hierarcheom}
 \partial_{t} \sigma^{(\kappa)}_{j_{1}..j_{\kappa}}(t) &=&  
 - i \left[ H_{\text{DQD}} , \sigma^{(\kappa)}_{j_{1}..j_{\kappa}}(t) \right] 
 - \sum_{\lambda\in\{1..\kappa\}} \omega_{K_{\lambda},p_{\lambda}}^{s_{\lambda}} \sigma^{(\kappa)}_{j_{1}..j_{\kappa}}(t) \\
 &&\hspace{-1cm} + \sum_{\lambda\in\{1..\kappa\}} (-1)^{\kappa-\lambda} 
 \eta_{K_{\lambda},m_{\lambda}n_{\lambda},p_{\lambda}}^{s_{\lambda}} d_{m_{\lambda}}^{s_{\lambda}} \sigma^{(\kappa-1)}_{j_{1}..j_{\kappa}/j_{\lambda}}(t) 
 + \sum_{\lambda\in\{1..\kappa\}} (-1)^{\lambda} \eta_{K_{\lambda},m_{\lambda}n_{\lambda},p_{\lambda}}^{\overline{s}_{\lambda},*}  \sigma^{(\kappa-1)}_{j_{1}..j_{\kappa}/j_{\lambda}}(t) d_{m_{\lambda}}^{s_{\lambda}} \nonumber \\
 &&\hspace{-1cm} - \sum_{j_{\kappa+1},n_{\kappa+1}} 
 \left( d_{n_{\kappa+1}}^{\overline{s}_{\kappa+1}} 
 \sigma^{(\kappa+1)}_{j_{1}..j_{\kappa}j_{\kappa+1}}(t) - (-1)^{\kappa} 
  \sigma^{(\kappa+1)}_{j_{1}..j_{\kappa}j_{\kappa+1}}(t) d_{n_{\kappa+1}}^{\overline{s}_{\kappa+1}} \right), \nonumber
\end{eqnarray} 
involve only the auxiliary operators $\sigma^{(\kappa+1)}_{j_{1}..j_{\kappa+1}}(t)$. The reduced density matrix enters 
this hierarchy of equations of motion at the $0$th tier as $\sigma^{(0)}(t)=\sigma(t)$. 
Truncation of the hierarchy at 
the $\kappa$th tier corresponds to an expansion in the hybridization versus the temperature in the leads 
(cf.\ the discussion given in Ref.\ \onlinecite{Hartle2013b}, where, in addition, further details on the numerical 
evaluation of the hierarchy of equations of motion (\ref{hierarcheom}) can be found). Note that the latter statement is 
strictly speaking only true in the strong coupling limit, $U\gg\Gamma_{K,mn}$. In the non-interacting limit ($U=0$) 
it has been found \cite{Jin2008,Jin2010} that the hierarchy (\ref{hierarcheom}) terminates already 
at the second tier.

\subsection{Born-Markov master equation approach} 
\label{bornmarkovtheory}

The hierarchical equation of motion technique (cf.\ Sec.\ \ref{heomtheory}) allows us to obtain the dynamics of the system 
in a numerically exact and systematic way. In addition, we employ the Born-Markov master equation method. 
The comparison to the HQME results will facilitate a better understanding of the underlying physics. 

Born-Markov master equations are well established 
\cite{May02,Mitra04,Lehmann04,Harbola2006,Volkovich2008,Hartle09,Hartle2010,Hartle2010b}. Here the reduced 
density matrix $\sigma$ is determined by the equation of motion 
\begin{eqnarray}
\label{genfinalME}
\frac{\partial \sigma(t)}{\partial t} &=& -i \left[ 
H_{\text{DQD}} , \sigma(t) \right]  - \int_{0}^{t} \text{d}\tau\, 
\text{tr}_{\text{L+R}}\lbrace \left[ H_{\text{tun}} , \left[ \tilde{H}_{\text{tun}}(\tau), \sigma(t) 
\sigma_{\text{L}} \sigma_{\text{R}} \right] \right] \rbrace , 
\end{eqnarray}
where 
\begin{eqnarray}
\tilde{H}_{\text{tun}}(\tau) = 
\text{e}^{-i(H_{\text{DQD}}+H_{\text{L}}+H_{\text{R}})\tau} 
H_{\text{tun}} \text{e}^{i(H_{\text{DQD}}+H_{\text{L}}+H_{\text{R}})\tau}.
\end{eqnarray}
It can be derived from the Nakajima-Zwanzig equation \cite{Nakajima,Zwanzig}, employing a 
second-order expansion in the coupling $H_{\text{tun}}$ and the so-called Markov approximation. 
Solving Eq.\ (\ref{genfinalME}) constitutes a time-dependent Born-Markov scheme (t-BM).

Due to the approximations involved, the master equation (\ref{genfinalME}) describes a non-unitary 
time evolution of the reduced density matrix, which can result in unphysical negative populations \cite{Egorova03,Kulkarni2013}. 
This problem can be avoided by shifting the integration limit $t$ to $\infty$ and, at the same time, 
neglecting principal value terms that arise in the evaluation of the resulting integrals. This is a 
standard procedure and we refer to it as the standard Born-Markov scheme (s-BM). A comparison of the s-BM and 
t-BM schemes helps to elucidate the role of the principal value terms. These give rise to both an interaction-induced 
renormalization \cite{Wunsch2005}  
and renormalization effects due to the structure of the conduction band \cite{Hartle2013b}.  
As we will see, these renormalization effects, which are not captured in the s-BM approximation, 
have a direct influence on the coherence, which, in turn, also affects the population of the dots.

Finally, we remark that we evaluate the HQMEs and the BM master equations in the basis of the states 
that are localized on dots $a$ and $b$. 
This includes $\{\vert00\rangle,\vert a \rangle,\vert b\rangle,\vert11\rangle\}$, which stands for an 
empty system, one/no electron in dot $a$/$b$, one/no electron in dot $b$/$a$, and a doubly occupied DQD. 
If the Born-Markov equation (\ref{genfinalME}) is evaluated in the eigenbasis of the system Hamiltonian $H_\text{DQD}$, 
it is equivalent to the Redfield (or Bloch-Wangsness-Redfield) 
equations \cite{Wangsness1953,Redfield1965,Egorova03,May04,Timm08}. 
Note that neither the HQME (in particular our truncation scheme \cite{Hartle2013b}) 
nor the BM formalism depends on the choice of the basis.

\subsection{Observables of interest}

We characterize the dynamics of the double dot system by following the time evolution of the the inter-dot 
coherence $\sigma_{a,b}$ and the dot populations. The latter includes the population of the doubly occupied 
state $\sigma_{11,11}$ and the populations of dot a/b, $\sigma_{a/b,a/b}$. 
Since $\text{Tr}_{\text{DQD}}\left[\sigma\right]=1$, the population of the empty state is given 
by $\sigma_{00,00}=1-\sigma_{11,11}-\sigma_{a,a}-\sigma_{b,b}$. While the populations represent the probability 
to find the system in the corresponding state, the coherence $\sigma_{a,b}$ describes the entanglement of the dots 
generated in coherent tunnelling processes between the dots themselves and the leads. If the inter-dot 
coupling is strong, the eigenstates of the double dot system are well separated in energy. 
The populations and the inter-dot coherence are, therefore, very similar. Their dynamics becomes less trivial if 
the coupling between the dots is small compared to the coupling to the electrodes. However, in the limit where the 
dots are not coupled, $\alpha\rightarrow0$, the coherence $\sigma_{a,b}$ vanishes (as for a Kondo impurity).

In experiment, the current that is flowing through the system (if a bias voltage is applied) is less 
directly affected by the dynamics of the system, because its detection requires millions of tunneling electrons. 
In contrast, the populations can be read out more efficiently and for each quantum dot independently using 
single-electron transistors or quantum point contacts \cite{Fujisawa2006,WangPetta2013,House2014}. 
Thus, we restrict our discussion in the following to the density matrix of the double dot structure.

\section{Results}
\label{results}

We investigate the dynamics of the quantum dot array that is depicted in Fig.\ \ref{graphicsl}. 
To this end, we focus on two complementary realiziations: a serial coupling configuration, 
where the two dots are connected in series, and a branched configuration, where only one of the 
dots is connected to the electrodes.  
These realizations are referred to as models SERIAL and BRANCHED in the following. 
The respective parameters can be found in Tab.\ \ref{parameters}.

We focus on coherent dynamics between the quantum dots and, therefore, on the parameter 
regime where the inter-dot coupling $\alpha$ is much weaker than the dot-lead coupling $\nu$. 
Note that for $\alpha=0$ the inter-dot coherence vanishes and that for a strong inter-dot coupling, 
the dynamics is governed by the eigenstates of the DQD. Only recently, we have given a detailed study of 
the steady-state properties of the systems SERIAL and BRANCHED (cf.\ Ref.\ \onlinecite{Hartle2013b}). 
We focused on decoherence phenomena and a lead-induced (RKKY-like) inter-state/dot coupling. 
Note that similar realizations of the spinless Anderson model have been considered both in a 
number of theoretical \cite{Gefen2002,Kubala2002,Cohen2007,Segal2012,Hartle2012} 
and experimental studies \cite{Holleitner2001,Holleitner2002,Iye2001,Wegscheider2007,Osorio2007,Nilsson2010}. 
These models have also been used to describe (linear or branched) nanoscale/molecular 
conductors \cite{Solomon2008b,Brisker2008,Hartle2012}.

\begin{table}
\begin{center}
\begin{tabular}{|*{12}{ccc|}}
\hline \hline 
&model &&& $\epsilon_{a}$&&&$\epsilon_{b}$&&&$\alpha$&&&$U$&&& $\nu_{\text{L},a}$&&& $\nu_{\text{L},b}$&&& $\nu_{\text{R},a}$&&&$\nu_{\text{R},b}$&&&
$\gamma$&\\ \hline 
& SERIAL &&&$\epsilon_{0}$&&&$\epsilon_{0}$&&&0.0005&&&0.5&&&$\nu$&&&0&&&0&&&$\nu$&&&2&\\
& BRANCHED &&&$\epsilon_{0}$&&&$\epsilon_{0}$&&&0.0005&&&0.5&&&$\nu$&&&0&&&$\nu$&&&0&&&2&\\
\hline \hline
\end{tabular}
\end{center}
\caption{\label{parameters}
Parameters of models SERIAL and BRANCHED, which represent a serial and a branched configuration of the 
double quantum dot system that is shown in Fig.\ \ref{graphicsl}, respectively. Energy values are given 
in eV. The dot-lead coupling parameter $\nu$ is set to $60$\,meV, 
corresponding to $\Gamma=2\pi\nu^{2}/\gamma\approx11$\,meV, and the level energy 
$\epsilon_{0}$ to $-150$\,meV. The temperature of the electrodes $T$ is $300$\,K. 
The width of the respective conduction bands $\gamma$ is set to $2$\,eV. 
Note that these parameters reflect typical 
experimental values \cite{Holleitner2001,Wegscheider2007,Osorio2007,Nilsson2010} 
with respect to the temperature scale $k_{\text{B}}T\approx25$\,meV used in our numerical calculations. 
}
\end{table}

We start to follow the dynamics of the system from two different initial states. 
The first describes a situation where both dots are unpopulated and uncorrelated 
(\emph{i.e.} $\sigma_{00,00}(t=0)=1$ while all other elements of the reduced density matrix are zero). 
The second differs from the first one by an electron in dot $a$, that is we set $\sigma_{a,a}(t=0)=1$ 
(and again all other elements to zero). These initial states are complementary in the sense that they 
describe a symmetric and an asymmetric distribution of charge in the DQD system and allow us to represent 
the full complexity of the underlying physics. They can be experimentally realized, for example,  
by a gate-voltage and/or a dot-lead coupling quench. In addition, we focus on systems that carry a single 
electron on average, \emph{i.e.} $\epsilon_{a/b}<\mu_{\text{L/R}}<\epsilon_{a/b}+U$ (see Fig.\ \ref{graphicsl}(b)). 
Systems with a different level structure ($\mu_{\text{L/R}}<\epsilon_{a/b},\epsilon_{a/b}+U$ or 
$\mu_{\text{L/R}}>\epsilon_{a/b},\epsilon_{a/b}+U$) do not exhibit the slow relaxation dynamics we are 
interested in (data not shown). It was also not observed at higher bias 
voltages $\Phi>2\text{min}(\vert\epsilon_{a/b}\vert,\vert\epsilon_{a/b}+U\vert)$. 
Throughout this work, we assume a Lorentzian form of the tunneling efficencies (which are 
defined by Eq.\ (\ref{tuneff})) 
\begin{eqnarray} 
\label{lorbands}
 \Gamma_{K,mn}(\epsilon) &=& 2\pi\sum_{k\in K} V_{mk}^{*} V_{nk}\delta(\epsilon-\epsilon_{k}) 
 \,=\, 2\pi \nu_{K,m} \nu_{K,n} \frac{\gamma}{(\epsilon-\mu_{K})^{2}+\gamma^{2}}.  
\end{eqnarray}
This is not a crucial assumption for the following but beneficial for the 
numerical evaluation of the HQME \cite{Jin2008,Hartle2013b}.

\subsection{Coherent charge oscillations and interaction-induced renormalization at zero bias}
\label{nonmarkovian}

We begin our discussion with the dynamics of the unbiased systems. The effect of a non-zero bias 
voltage will be considered in Sec.\ \ref{feedback}. This procedure allows us to distinguish equilibrium 
and nonequilibrium effects. It also elucidates qualitative differences between the Born-Markov schemes, 
the HQME approach and a truncation of the HQME at the first tier. Such differences are interesting not only 
from a methodological point of view but enable us to elucidate the underlying physical mechanisms 
that are at work in these systems.

\begin{figure}
\begin{center}
\includegraphics[width=\textwidth]{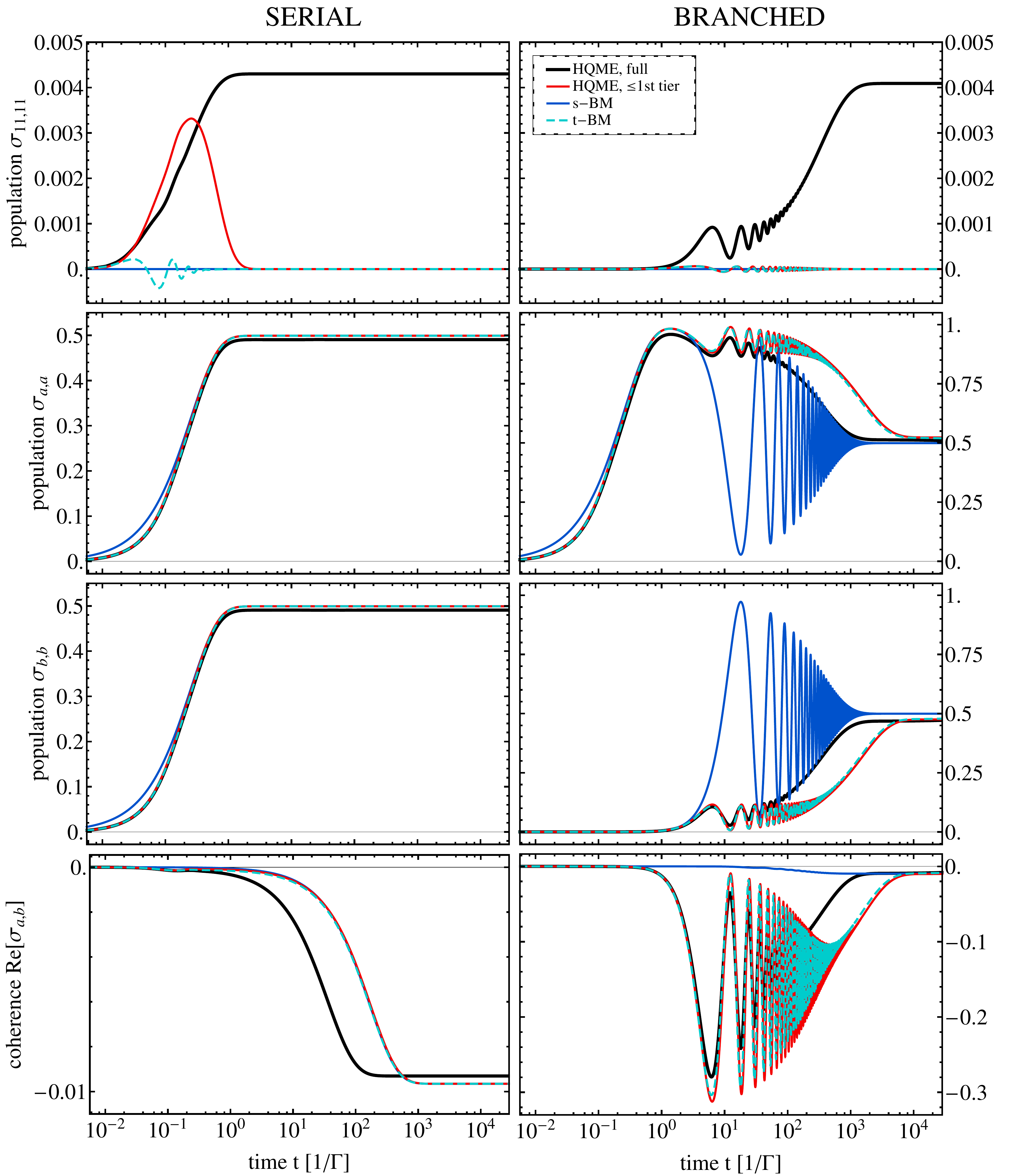}
\end{center}
\caption{(Color online) \label{NonMarkovianFig} 
Population of the doubly occupied state, the single-particle levels in dots $a$ and $b$ and the real part 
of the coherence $\sigma_{a,b}$ as functions of time, starting with the unpopulated system ($\sigma_{00,00}(t)=1$). 
The left and the right column show these functions for the unbiased systems SERIAL and BRANCHED, respectively. 
}
\end{figure}

\begin{figure}
\begin{center}
\includegraphics[width=\textwidth]{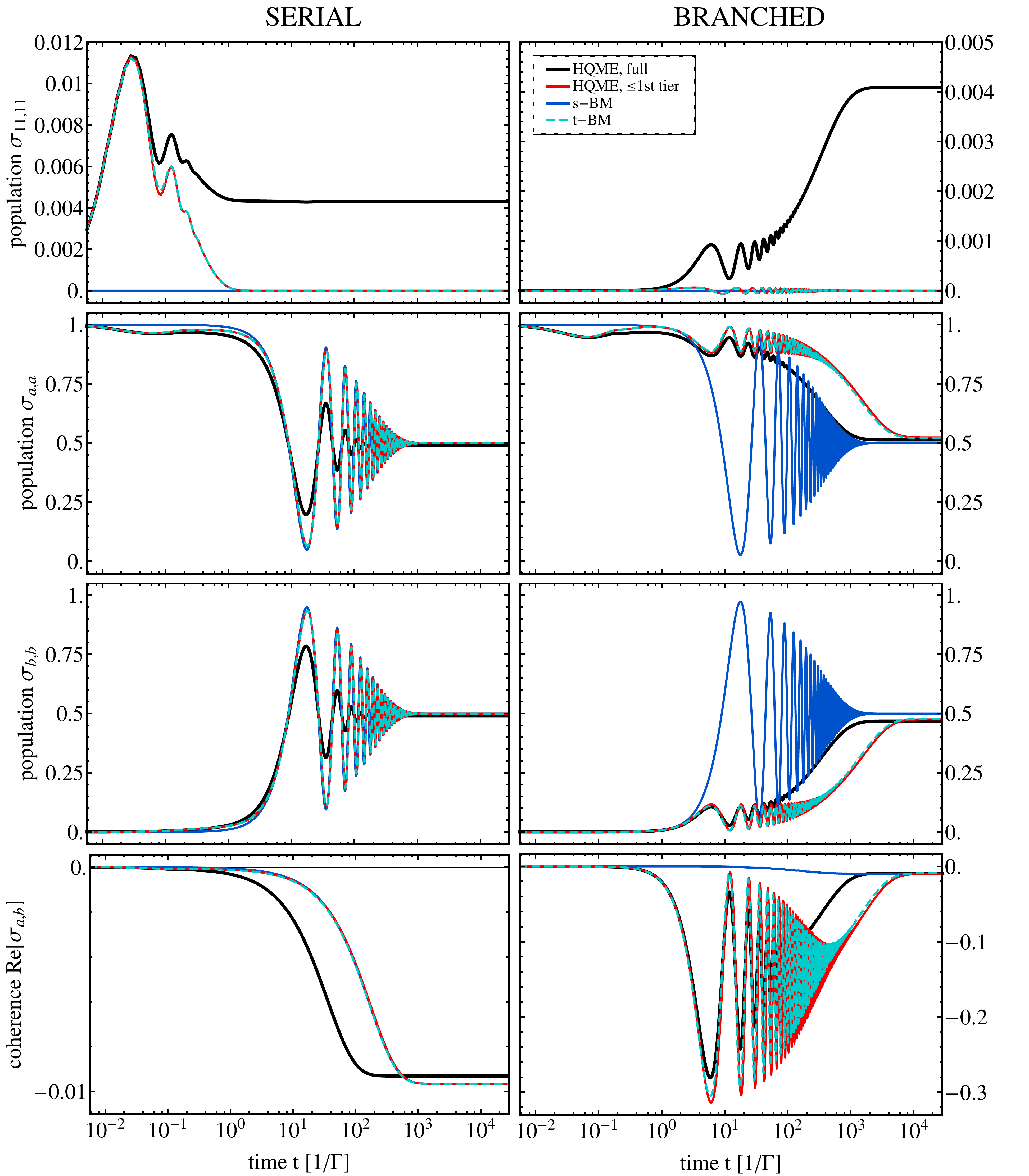}
\end{center}
\caption{(Color online) \label{NonMarkovianFig1010} 
Population of the doubly occupied state, the single-particle levels in dots $a$ and $b$ and the real part 
of the coherence $\sigma_{a,b}$ as functions of time, starting with an electron in dot $a$ ($\sigma_{a,a}(t)=1$). 
The left and the right column show these functions for the unbiased systems SERIAL and BRANCHED, respectively. 
}
\end{figure}

It turns out that the dynamics of systems SERIAL and BRANCHED can be fully characterized by four elements 
of the reduced density matrix: the population of the doubly occupied state, the population of the 
single-particle levels in dots $a$ and $b$ and the real part of the coherence $\sigma_{a,b}$. 
These quantities are depicted in Figs.\ \ref{NonMarkovianFig} and \ref{NonMarkovianFig1010},  
where the top rows show the population of the doubly occupied state, the second and third rows the 
single-particle population of dots $a$ and $b$ and the bottom rows the real part of the 
coherence $\sigma_{a,b}$. Fig.\ \ref{NonMarkovianFig} depicts the dynamics of systems SERIAL and BRANCHED 
starting from the symmetric ($\sigma_{00,00}(0)=1$) and Fig.\ \ref{NonMarkovianFig1010} from the asymmetric 
initial state ($\sigma_{a,a}(0)=1$), where the left columns refer to system SERIAL while the right 
ones depict the behavior of system BRANCHED. The exact result, which has been obtained by solving the 
full HQMEs, is depicted by solid black lines. It is compared to three approximate results, where the 
HQMEs are truncated at the first tier (solid red lines) and where the standard (s-BM) and the 
time-dependent Born-Markov scheme (t-BM) have been used (solid blue and dashed turquoise lines, respectively).

We consider first the exact dynamics of model SERIAL, starting from the unpopulated system 
(black lines on the left of Fig.\ \ref{NonMarkovianFig}). The corresponding populations show a decay of the 
initial state to a state, where the two dots are equally occupied and host, on average, a single electron. 
This behavior is typical for a double dot structure where the single-particle levels $\epsilon_{a/b}$ are 
located below and the states associated with double occupation (at energies $\epsilon_{a/b}+U$) 
above the chemical potentials in the leads. It is dominated by resonant tunneling processes from the electrodes 
onto the dots and, therefore, occurs on time scales $\sim1/\Gamma=1/\Gamma_{K,mm}(\mu_{K})$. A very similar behavior 
can be observed in the dynamics of a Kondo impurity \cite{Cohen2013}.

For junction BRANCHED (black lines on the right of Fig.\ \ref{NonMarkovianFig}), the situation is more complex. 
Initially, (\emph{i.e.}\ on time scales $1/\Gamma$), the population of dot $a$ increases to values that are close 
to one, while dot $b$ remains almost unpopulated. This is related to both the position of the energy levels 
($\epsilon_{0}\ll\mu_{\text{L/R}}$) and the geometry of the device, where tunneling onto dot $b$ is only possible 
via dot $a$. These tunneling processes involve a coherent charge transfer from dot $a$ to dot $b$, 
which is facilitated by the weak inter-dot coupling $\alpha$. Therefore, dot $b$ is populated on much longer 
time scales, \emph{i.e.}\ about $\pi\Gamma/\alpha\approx100$ longer than the time scale to populate dot $a$. 
As the system approaches the steady state regime, the populations of the two dots evolve to $1/2$, 
reflecting the fact that tunneling on and off the dots occurs with the same probability.

\begin{figure}
\begin{center}
\includegraphics[width=\textwidth]{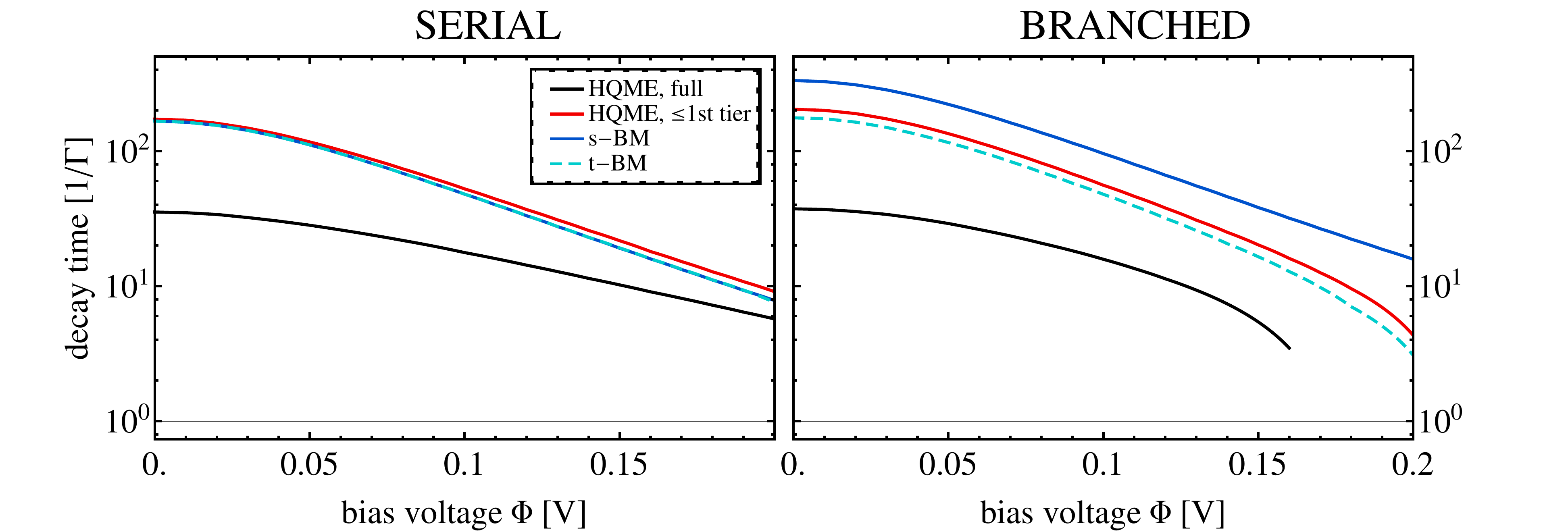}
\end{center}
\caption{(Color online) \label{decaytimes} 
Decay times of the coherent charge oscillations in junction SERIAL (left plot) and 
junction BRANCHED (right plot) as a function of the applied bias voltage, starting from the 
asymmetric initial state $\sigma_{a,a}(t=0)=1$ (which we used, because 
coherent charge oscillations are quenched in the SERIAL configuration if the 
symmetric initial state is used, cf.\ the discussion of Figs.\ \ref{NonMarkovianFig} and \ref{NonMarkovianFig1010}). 
To this end, we fitted the oscillation amplitude in $\sigma_{b,b}(t)$ to an exponential decay. 
}
\end{figure}

In addition, junction BRANCHED exhibits oscillations in the population of the two dots on 
intermediate time scales, $\sim1/\Gamma$ to $\sim10^{3}/\Gamma$. These oscillations reflect coherent 
charge transfer processes between the two dots \footnote{Note that a detailed study of such oscillations 
in transport through non-interacting quantum dots has been given, e.g., by Taranko \emph{et al.} \cite{Taranko2012}.}. 
The period of these oscillations is determined by the energy difference of the eigenstates and will be discussed 
in more detail below (see Eq.\ (\ref{frequcohosc})). 
Their coherent nature is underlined by a pronounced 
real part of the inter-dot coherence $\sigma_{a,b}$ (cf.\ the lower right plot of Fig.\ \ref{NonMarkovianFig}). 
The origin of these oscillations is an asymmetry in the dot population. Naturally, they become 
suppressed in the steady state regime because the populations of the two dots become very similar. 
In the steady state regime, the presence of dot $b$ thus reduces to an electrostatic effect 
(cf.\ our findings in Ref.\ \onlinecite{Hartle2013b}). The suppression of the coherent charge oscillations 
can be fitted to an exponential decay. The corresponding decay time is given 
in Fig.\ \ref{decaytimes} (see the zero bias value of the right plot) and is of the order of $\sim10/\Gamma$.

Coherent charge oscillations are also observed in the dynamics of junction SERIAL if the initial 
charge distribution is asymmetric. This can be seen by the black lines on the left of Fig.\ \ref{NonMarkovianFig1010}, 
where we depict the dynamics starting from an initially asymmetric population of the dots ($\sigma_{a,a}(t=0)=1$). 
The corresponding decay time is similar to the one in junction BRANCHED, \emph{i.e.}\ $\sim10/\Gamma$ 
(see the value at zero bias in the left plot of Fig.\ \ref{decaytimes}). For junction BRANCHED, 
the influence of such an asymmetry is less pronounced (compare the black lines on the right of 
Figs.\ \ref{NonMarkovianFig} and\ref{NonMarkovianFig1010}), as it develops naturally from the geometry of the device. 
Similar effects are observed if the two quantum dots are coupled asymmetrically to the electrodes (data not shown).  
Overall, however, we do not observe any dependence of the steady state on the initial state, even in the biased 
scenarios discussed in Sec.\ \ref{feedback}.

\begin{figure}
\begin{center}
\includegraphics[width=\textwidth]{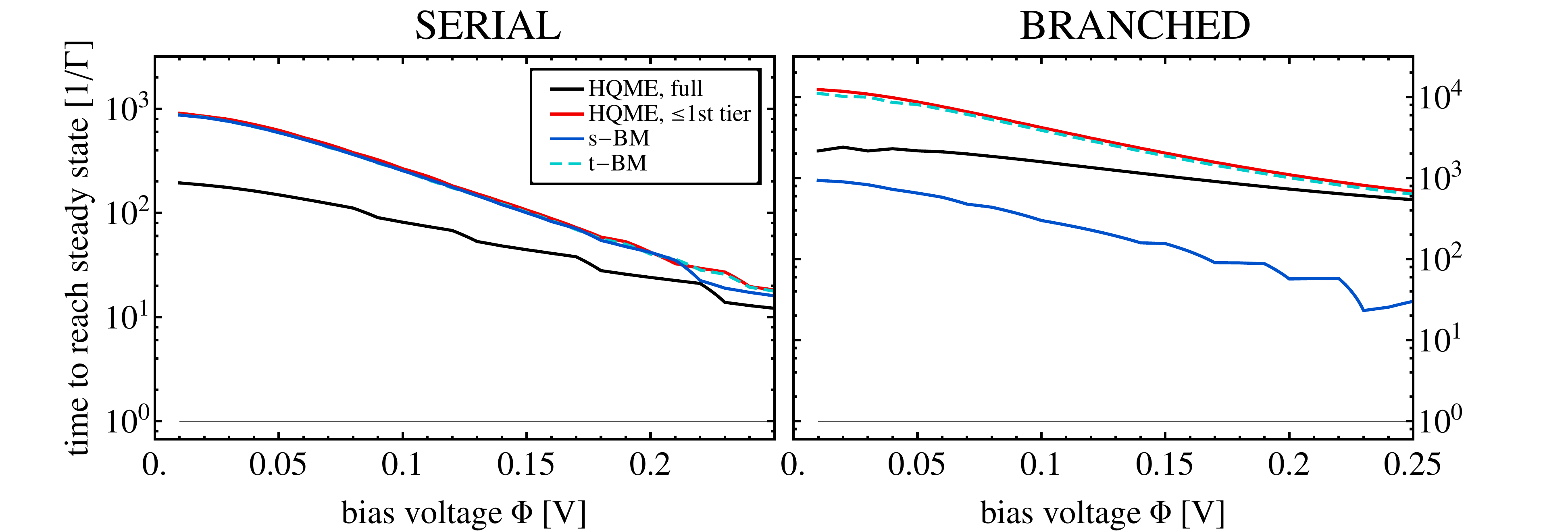}
\end{center}
\caption{(Color online) \label{timescales} 
Time scale to reach the steady state in junction SERIAL (left plot) and junction BRANCHED (right plot) 
as a function of the applied bias voltage, starting from the symmetric initial state $\sigma_{00,00}(t=0)=1$ 
(which we used to avoid ambiguities due to the presence of coherent charge oscillations in the SERIAL configuration). 
To determine this scale, we use the time where the real part of the coherence deviates 
0.5\% from the steady state value. Note that the oscillatory behavior originates from 
dynamical phases and is, therefore, most pronounced when the steady state is reached 
on short time scales. 
}
\end{figure}

Further insights can be gained by comparing the exact result with the approximate ones. 
For example, a comparison of the black and the red lines elucidates the role of higher order processes. 
They increase the probability for electron exchange processes with the leads and, therefore, result in a 
quenching of coherent charge oscillations and a faster build-up of the steady state (see, for example, the 
dot populations shown in the two middle panels of Fig.\ \ref{NonMarkovianFig1010}). The time scale where 
the systems reach the steady state are quantified in Fig.\ \ref{timescales}. There, it can be seen that 
higher-order processes reduce the time scale to reach the steady state by almost an order of magnitude. 
This may not be surprising for systems that are operated in the non-resonant regime, 
that is for $\epsilon_{a/b}<\mu_{\text{L/R}}<\epsilon_{a/b}+U$ where resonant processes are suppressed 
such that non-resonant processes become important.

Differences between the red and the turquoise lines can also be understood in terms of higher-order processes, 
considering that the Markov approximation (without the shift of the integration limit) represents a more restrictive 
expansion to $\mathcal{O}(\Gamma)$. Thus, a pronounced intermediate population of the doubly occupied state appears in 
the red but not in the blue line (see top left plot of Fig.\ \ref{NonMarkovianFig}). Thereby, the HQME result appears to 
be more consistent, as t-BM gives unphysical negative populations (e.g.\ of the doubly occupied state).

The real part of the coherence (see bottom panels of Figs.\ \ref{NonMarkovianFig} and \ref{NonMarkovianFig1010}) develops 
on rather long time scales. This behavior is seen in both the HQME and the BM results, where the latter facilitate a 
direct access to the underlying physics. Using BM theory, the equation of motion of the coherence involves terms that 
involve the decay rates $\Gamma f(\overline{\epsilon}+U)$ and $\Gamma (1-f(\overline{\epsilon}))$. For the parameters 
considered, these rates are much smaller than the bare hybridization $\Gamma$, resulting in resonant dynamics on 
time scales $(1-f_{\text{L/R}}(\epsilon_{0}))^{-1}\approx300$ times the inverse of the hybridization strength $1/\Gamma$. 
Note that a non-zero value of the real part of the coherence signals a different population of the eigenstates of the DQD system. 
Considering the temperature in the leads and the energy difference of the eigenstates, which can be estimated 
by $2\alpha$\footnote{or, more rigorously, by the denominator of Eq.\ (\ref{frequcohosc})}, such a population difference 
is to be expected in the steady state at zero bias.

Another intriguing effect emerges from the comparison of the blue and the red/turquoise lines. 
This includes, for example, a reduction of the oscillation period by a factor of $\approx1/3$ 
in junction BRANCHED, which is visible, for example, in the two middle plots on the right hand 
side of Figs.\ \ref{NonMarkovianFig} and \ref{NonMarkovianFig1010}. These results can be 
qualitatively and quantitatively explained by the interaction-induced renormalization of energy levels, 
which has been outlined first by Wunsch \emph{et al.} \cite{Wunsch2005} in the context of 
double quantum dots and by Braun et al. for spin-valve setups \cite{Braun2004}. This renormalization 
is a combined effect of the local electron-electron interactions $U$ and the coupling of the dots to 
the electrodes and occurs not only for structured but also for flat conduction bands. 
For the systems of interest here, these renormalizations are given by: 
\begin{eqnarray}
\label{intindrenorm}
 \Delta \epsilon_{a/b,\text{L/R}} = \phi(\epsilon_{a/b},\mu_{\text{L/R}}) - \phi(\epsilon_{a/b}+U,\mu_{\text{L/R}}), 
\end{eqnarray}
with 
\begin{eqnarray}
 \phi(x,\mu) = 
 \frac{\Gamma}{2\pi}\text{Re}\left[ \Psi\left( \frac{1}{2} + \frac{i (x-\mu)}{2\pi k_{\text{B}}T} \right) \right],
\end{eqnarray}
and $\Psi(x)$ is digamma function\footnote{Eq.\ (\ref{intindrenorm}) 
can be straightforwardly derived from Eq.\ (9) of Ref.\ \onlinecite{Wunsch2005}, 
disregarding the spin contributions, that is using $U'=0$ and dividing by a factor of $2$.}. 
From the above formula, we can directly infer the aforementioned 
reduction of the oscillation period, which is given by 
\begin{eqnarray}
\label{frequcohosc}
 2\pi / \left(\sqrt{4\alpha^{2} + 
 \left( \sum_{K} \Delta \epsilon_{a,K} - \sum_{K} \Delta \epsilon_{b,K}\right)^{2} } \right).
\end{eqnarray}
We find (data not shown) that this renormalization is not present at the charge-symmetric point 
since $\phi(\epsilon_{0}+U,0)=\phi(-\epsilon_{0},0)=\phi(\epsilon_{0},0)$ \cite{Splettstoesser2012}. 
Moreover, it does not appear in junction SERIAL, since both levels are shifted in the same way 
at zero bias, \emph{i.e.}\ $\Delta\epsilon_{a,\text{L}}=\Delta\epsilon_{b,\text{R}}$. At this point, 
it should be noted that the interaction-induced renormalization is already active at times $\sim1/\Gamma$. 
For later reference, we also remark that the bias dependence of $\Delta \epsilon_{a/b,\text{L/R}}$ 
leads to additional shifts of the oscillation period, which are of the order of 10\% for the parameters 
considered in this work.

We conclude this section pointing out the different behavior of the s-BM scheme in more detail. 
For the branched system, for example, the s-BM approach gives very different results for the time scale 
to reach the steady state and the decay time of the coherent charge oscillations 
(cf.\ the right plots of Figs.\ \ref{decaytimes} and \ref{timescales}). This is of course related to 
the fact that the s-BM scheme misses the interaction-induced renormalizations (\ref{intindrenorm}). 
For the same reason, the real part of the coherence that is obtained by the s-BM scheme 
does not develop the pronounced values that are obtained by the t-BM and the HQME methods 
(cf.\ the lower right plots of Figs.\ \ref{NonMarkovianFig} and \ref{NonMarkovianFig1010}). 
Moreover, at short times $t\ll1/\Gamma$, the s-BM scheme exhibits an exponential scaling with time, 
while the HQME and t-BM give a power-law scaling, $\sim t^{2}$ (see, for example, the middle panels 
of Fig.\ \ref{NonMarkovianFig}). This behavior is due to the shift of the integration limit in 
Eq.\ (\ref{genfinalME}) and has been outlined before by Thoss \emph{et al.}\ \cite{Thoss01,Egorova03}.

\subsection{Interplay of inter-dot coherence and dot populations 
due to coherent nonequilibrium dynamics} 
\label{feedback}

\begin{figure}
\begin{center}
\includegraphics[width=\textwidth]{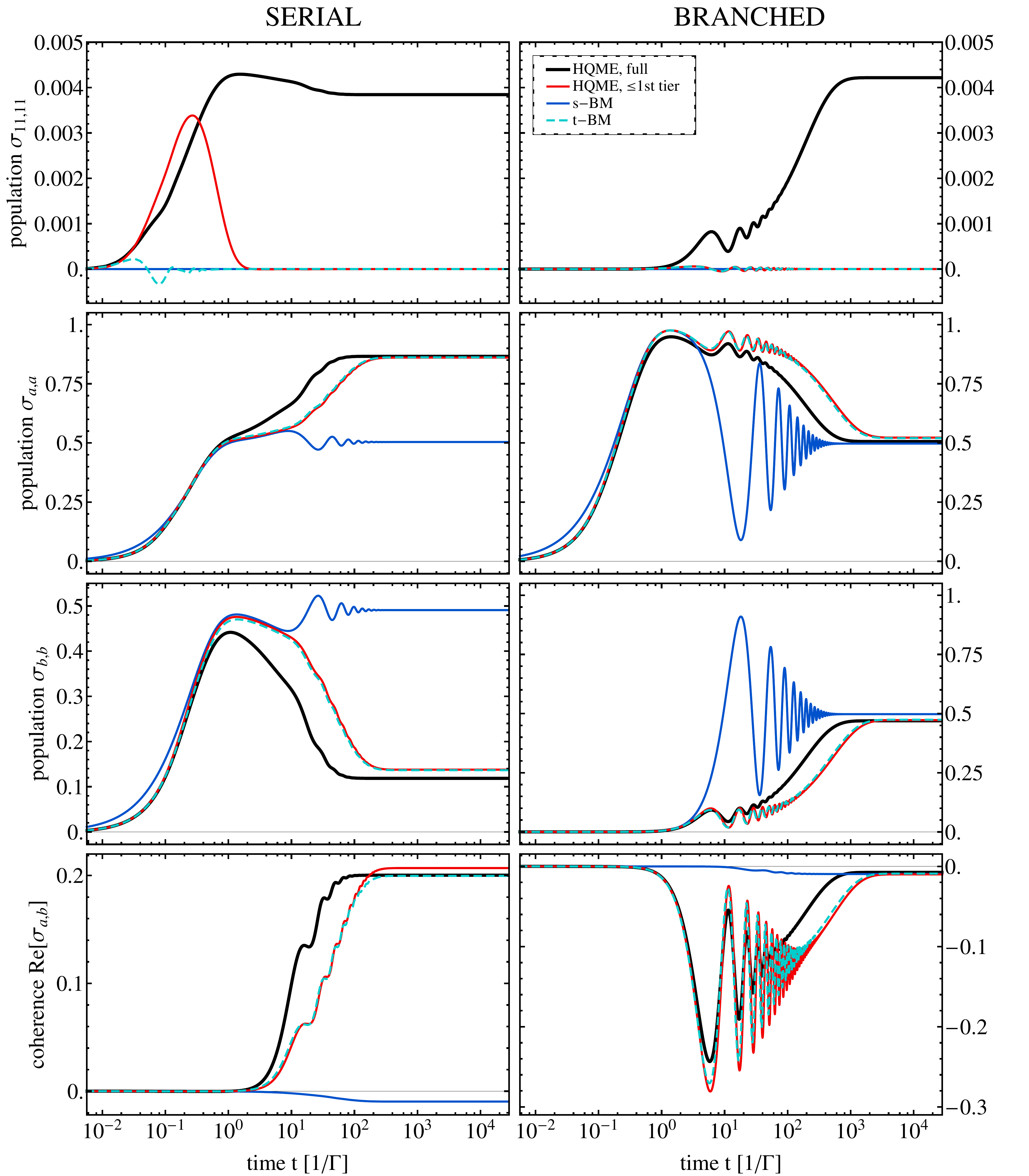}
\end{center}
\caption{(Color online) \label{NonequilibriumFig} 
Population of the doubly occupied state, the single-particle levels in dots $a$ and $b$ and the real part 
of the coherence $\sigma_{a,b}$ as functions of time, starting with the unpopulated system ($\sigma_{00,00}(t)=1$). 
The left and the right column show these functions for the systems SERIAL and BRANCHED, respectively, 
where a bias voltage of $\Phi=0.1$\,V is applied. 
}
\end{figure}

\begin{figure}
\begin{center}
\includegraphics[width=\textwidth]{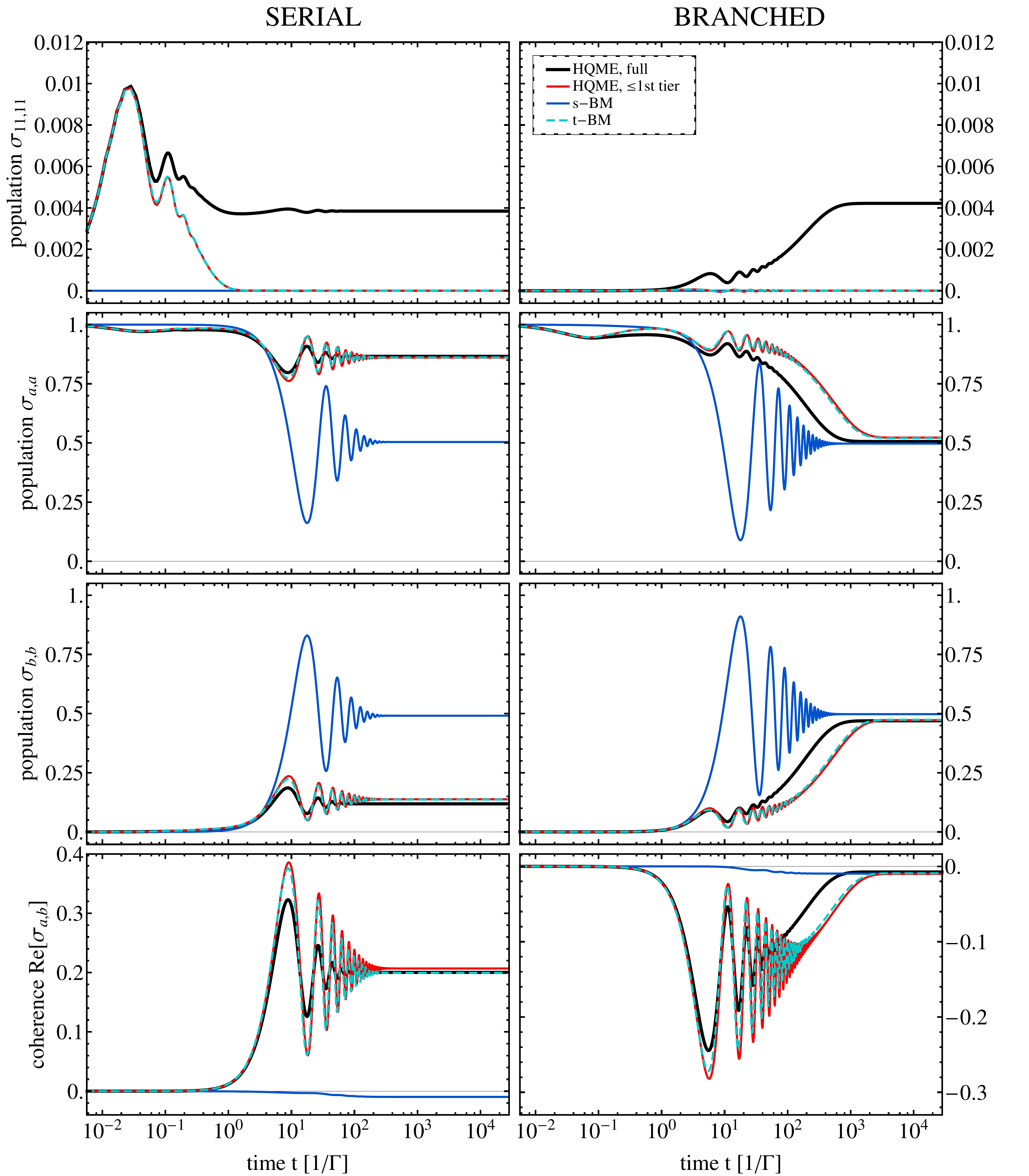}
\end{center}
\caption{(Color online) \label{NonequilibriumFig1010} 
Population of the doubly occupied state, the single-particle levels in dots $a$ and $b$ and the real part 
of the coherence $\sigma_{a,b}$ as functions of time, starting with an electron in dot $a$ ($\sigma_{a,a}(t)=1$). 
The left and the right column show these functions for the systems SERIAL and BRANCHED, respectively, 
where a bias voltage of $\Phi=0.1$\,V is applied. 
}
\end{figure}

In this section we study the dynamics of systems SERIAL and BRANCHED in the presence of a bias voltage. 
We restrict the discussion to the non-resonant transport regime and choose, accordingly, 
a low value ($\Phi=0.1$\,V) for the bias voltage such that the filled and empty states remain far 
from the chemical potential of either lead. At higher bias voltages, we do not observe the complex 
long-time behavior we are interested in. We characterize the nonequilibrium dynamics of the biased 
systems by the same quantities as the equilibrium dynamics of the unbiased ones. 
They are depicted in Figs.\ \ref{NonequilibriumFig} and \ref{NonequilibriumFig1010}, 
corresponding to an initially symmetric and asymmetric charge configuration, respectively.

At first sight, most of the dynamics is very similar to the one of the equilibrium case. 
The steady state is reached slightly faster in the presence of a bias voltage (cf.\ Fig.\ \ref{timescales}). 
Also, the coherent charge oscillations decay slightly faster (cf.\ Fig.\ \ref{decaytimes}). The main reason 
for this behavior is that the energy levels of the dots are closer to the chemical potential in the leads. 
The respective exponential scaling, which is observed once the bias voltage exceeds the thermal broadening in 
the two electrodes, \emph{i.e.}\ for $\Phi>0.05$\,V, is inherited from the bias dependence of the rates 
$\Gamma f(\overline{\epsilon}+U)$ and $\Gamma (1-f(\overline{\epsilon}))$ for resonant tunneling processes. 
There are, however, also a number of qualitative differences if a bias voltage is applied to 
systems SERIAL and BRANCHED.

The most pronounced response to an external bias voltage is observed in the SERIAL device. 
The real part of the inter-dot coherence $\sigma_{a,b}$, for example, acquires a different sign and its absolute 
value increases by more than order of magnitude to $\approx0.2$ (compare, for example, the bottom left plot of 
Figs.\ \ref{NonMarkovianFig} and \ref{NonequilibriumFig}). Moreover, the populations of the two quantum dots no 
longer evolve to the same value. The double dot structure still carries a single electron on average, but this 
electron is now more likely to be found in dot $a$ with a difference in the dot population that amounts 
to $\approx75$\% (cf.\ the dot populations shown on the left of Fig.\ \ref{NonequilibriumFig}). The corresponding 
time evolution develops on rather long time scales, that is $\sim10/\Gamma$ -- $\sim100/\Gamma$. This behavior is 
captured by the HQME and t-BM scheme but is missed by the s-BM approach. We can therefore relate it to the principal 
value terms that are included in the HQME and t-BM scheme but discarded in the s-BM approach. These terms include 
the interaction-induced renormalization, which we already pointed out in Sec.\ \ref{nonmarkovian}, and a renormalization 
due to the band width $\gamma$ \cite{Hartle2013b}. Since we observe qualitatively and quantitatively the same effects 
for different band widths $\gamma$ (where the coupling strength $\nu$ needs to be adjusted to give the same values 
for $\Gamma(\epsilon_{0})$), we attribute these effects to the interaction-induced 
renormalizations $\Delta\epsilon_{a/b,\text{L/R}}$. We continue to analyze this behavior in more detail.

\begin{figure}
\begin{tabular}{l}
\hspace{-0.5cm}(a) \\
\resizebox{\newwidth}{\newheight}{
\includegraphics{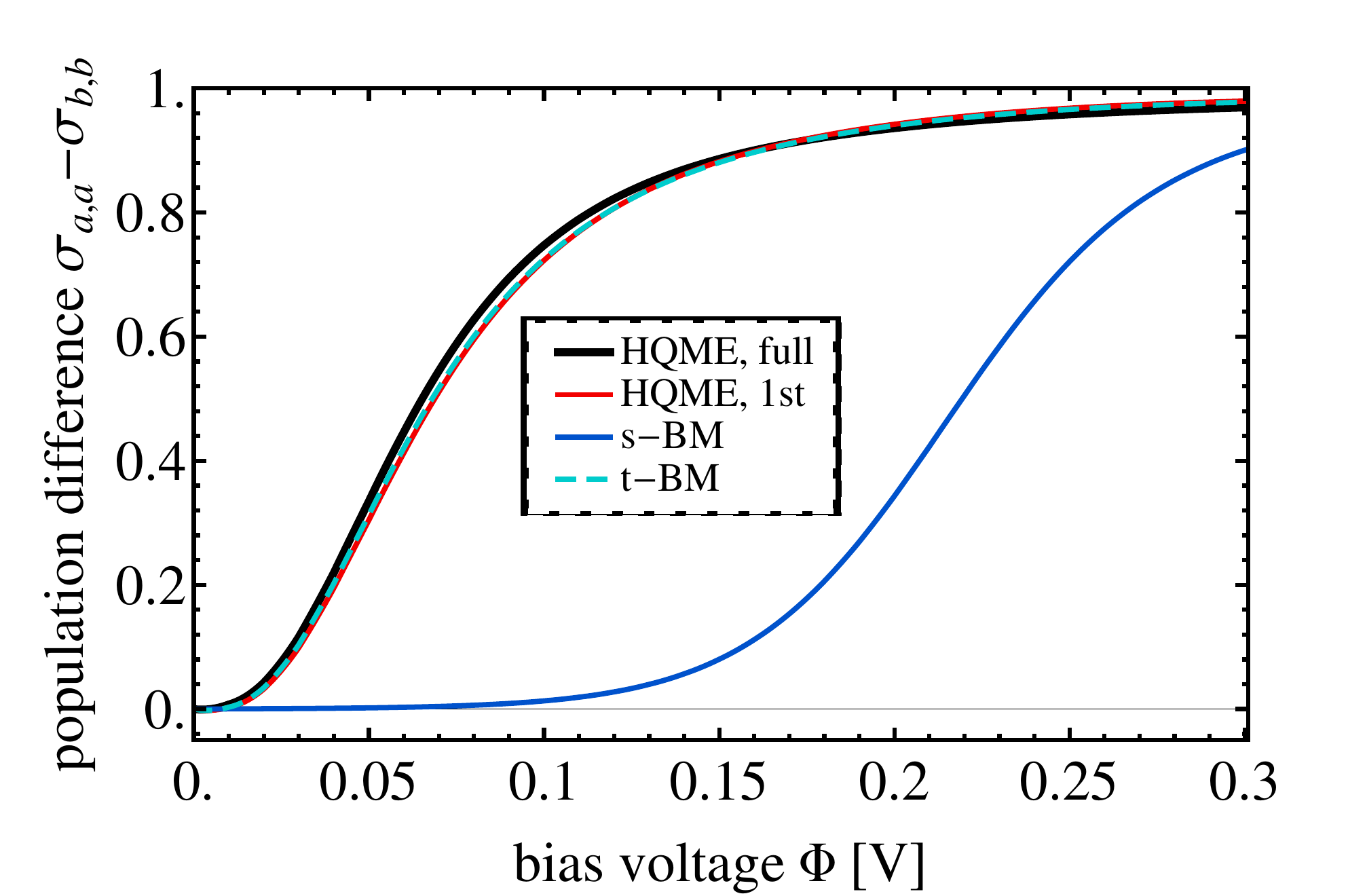}
}\\
\hspace{-0.5cm}(b) \\
\resizebox{\newwidth}{\newheight}{
\includegraphics{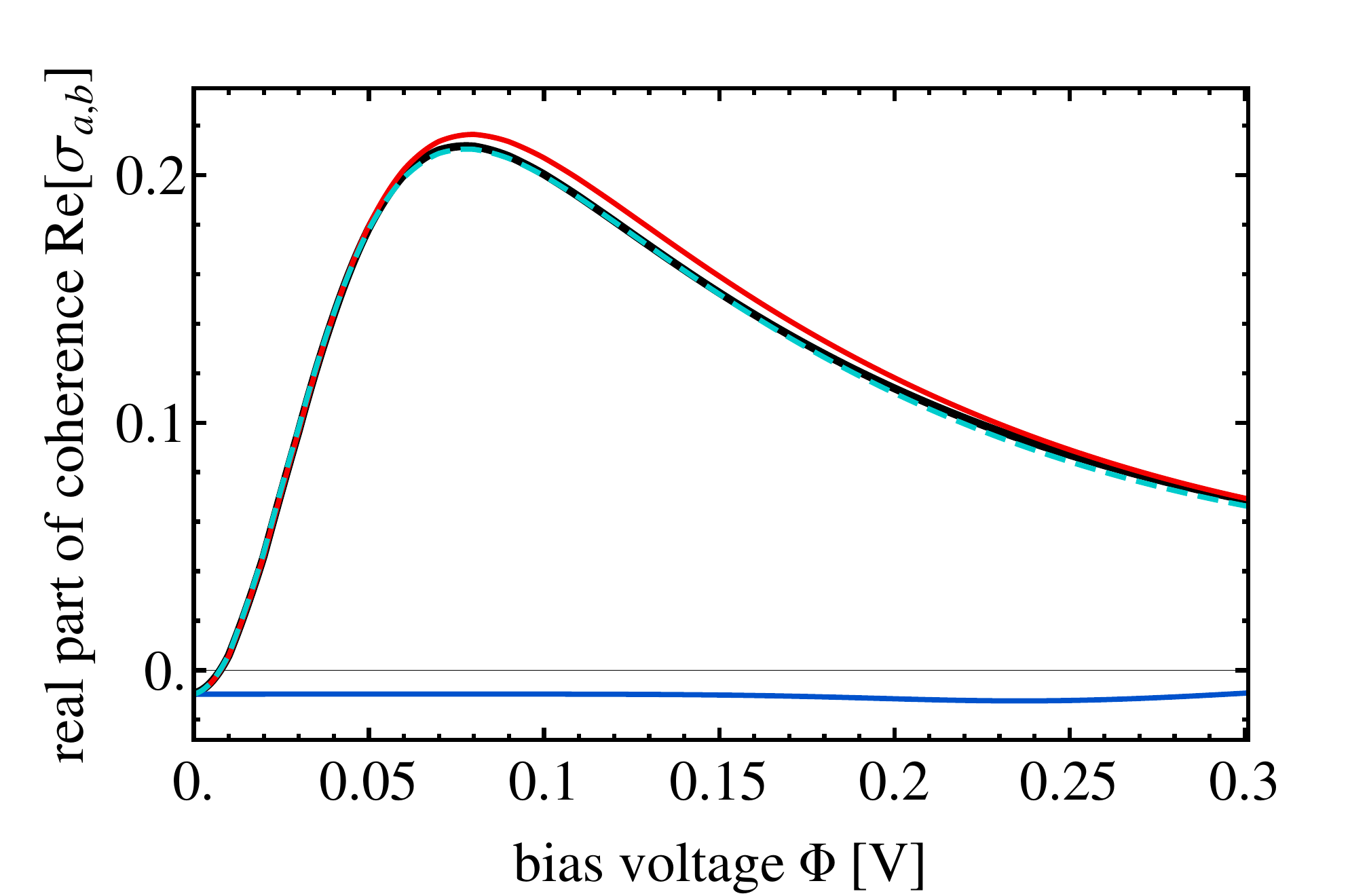}
}\\
\end{tabular}
\caption{(Color online) \label{biasdiff} 
Difference in the population of the dots $a$ and $b$ and the real part of the 
coherence $\sigma_{a,b}$ as a function of the bias voltage $\Phi$ applied to junction SERIAL. 
Note that the bias voltage is to be compared with the width of the transport resonances, 
which, in the present context, is given predominantly by the temperature scale, $k_{\text{B}}T\approx25$\,meV. 
}
\end{figure}

\begin{figure}
\begin{tabular}{l}
\hspace{-0.5cm}(a) \\
\resizebox{\newwidth}{\newheight}{
\includegraphics{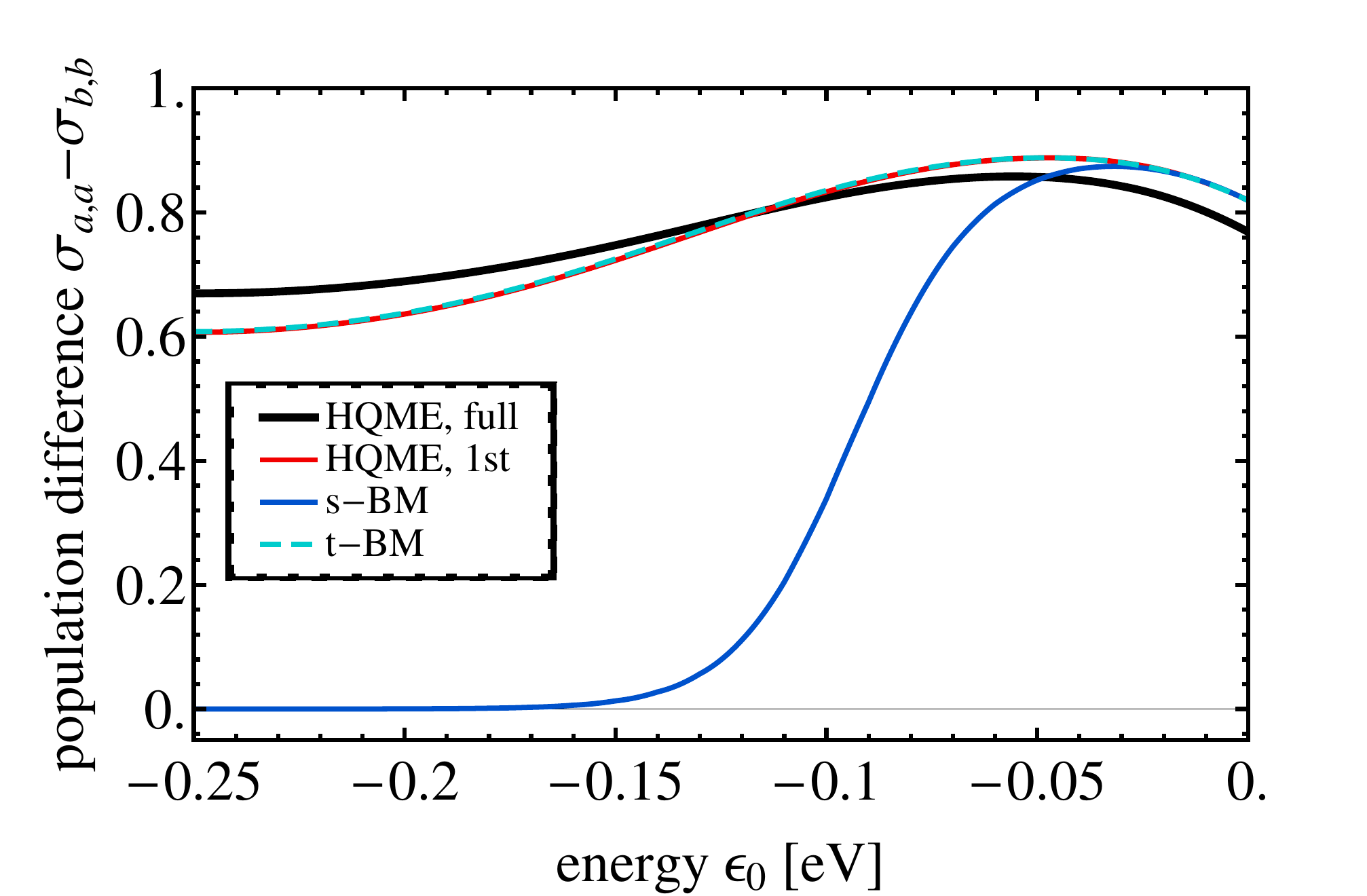}
}\\
\hspace{-0.5cm}(b) \\
\resizebox{\newwidth}{\newheight}{
\includegraphics{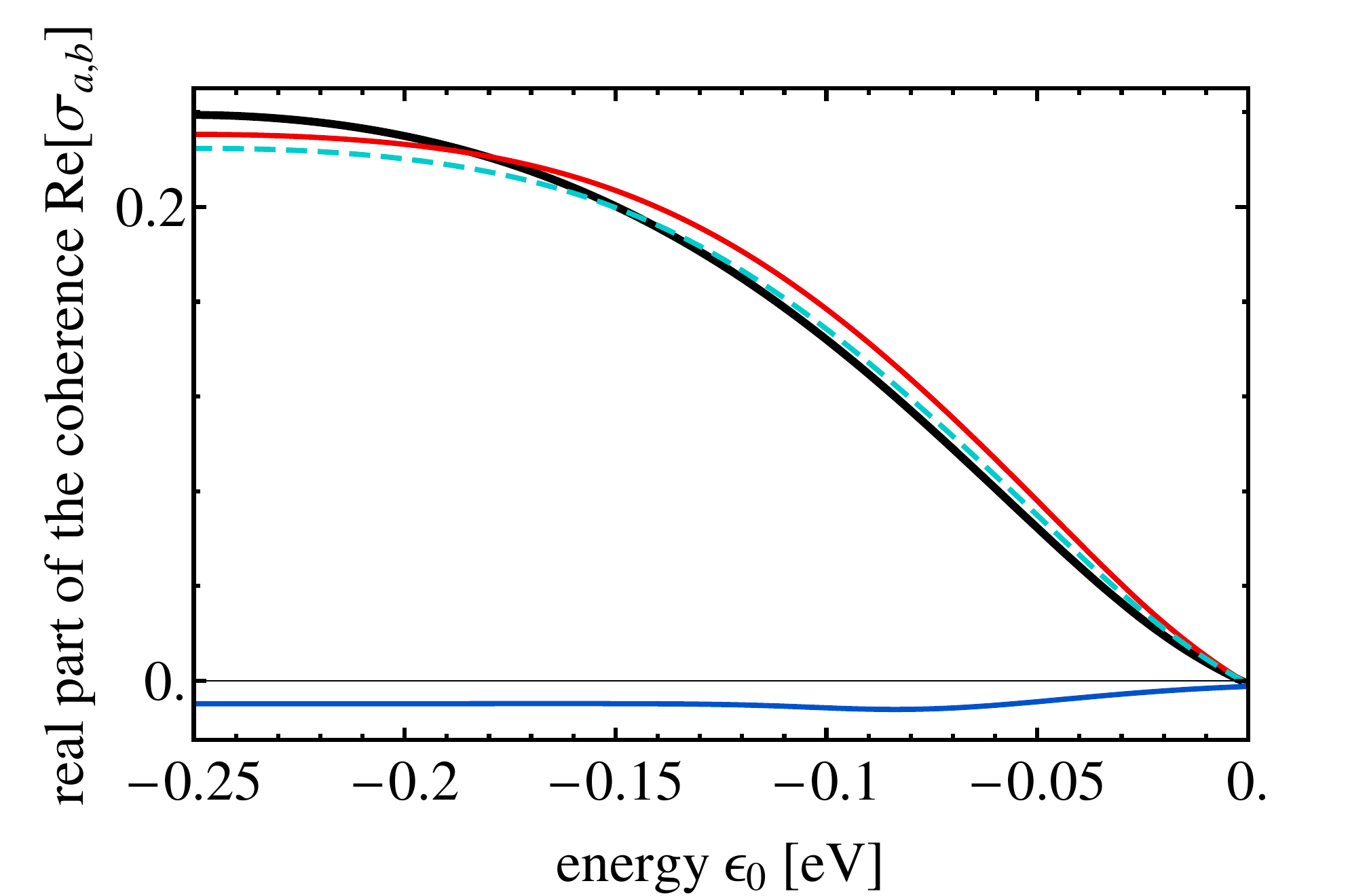}
}\\
\end{tabular}
\caption{(Color online)\label{diffpop} 
Difference in the population of the dots $a$ and $b$ and the corresponding real part of the 
coherence $\sigma_{a,b}$ as a function of the energy level position $\epsilon_{0}$ in junction SERIAL 
at bias voltage $\Phi=0.1$\,V. The scale of the level position $\epsilon_{0}$ is, 
similar to the bias voltage, determined by the temperature $k_{\text{B}}T\approx25$\,meV. 
}
\end{figure}

At first glance, it may not be surprising that, for positive bias voltages, the population of dot $a$ is higher 
than the one of dot $b$ (and vice versa for negative bias voltages). Since the inter-dot coupling $\alpha$ is much 
weaker than the coupling of the dots to the electrodes, the tunneling electrons are expected to get stuck at the 
inter-dot tunneling barrier. This can be seen in Figs.\ \ref{biasdiff}(a) and \ref{diffpop}(a), where the steady state 
population difference in system SERIAL is depicted as a function of the applied bias voltage and the level 
energy $\epsilon_{0}$, respectively. At the onset of the resonant transport regime, which corresponds 
to $\Phi\gtrsim2(\epsilon_{0}-k_{\text{B}}T)$ in Fig.\ \ref{biasdiff}(a) or 
to $\epsilon_{0}>-k_{\text{B}}T$ in Fig.\ \ref{diffpop}(a), the population difference is $\gtrsim0.8$. 
Here, the HQME and BM schemes yield very similar results.

The situation is different at lower bias voltages and / or closer to the charge-symmetric point. 
Due to the Pauli principle, the tunneling of an electron from one of the dots into the electrodes is 
suppressed by Fermi factors $(1-f_{\text{L/R}}(\epsilon_{0}))=3\cdot10^{-3}$, while the coherent 
transfer of electrons between the dots takes place on much shorter time 
scales $1/\alpha\ll(\Gamma(1-f_{\text{L/R}}(\epsilon_{0})))^{-1}$. Thus, an electron can be expected to 
tunnel many times between dots $a$ and $b$ before it enters one of the electrodes. While this suggests 
a population of the dots that is very similar, the HQME and t-BM data exhibit a pronounced bias-induced 
population difference, which can be orders of magnitude larger than the one obtained from the s-BM scheme 
(cf.\ Figs.\ \ref{biasdiff}(a) and \ref{diffpop}(a)).

\begin{figure}
\begin{tabular}{l}
\hspace{-0.5cm}(a) \\
\resizebox{\newwidth}{\newheight}{
\includegraphics{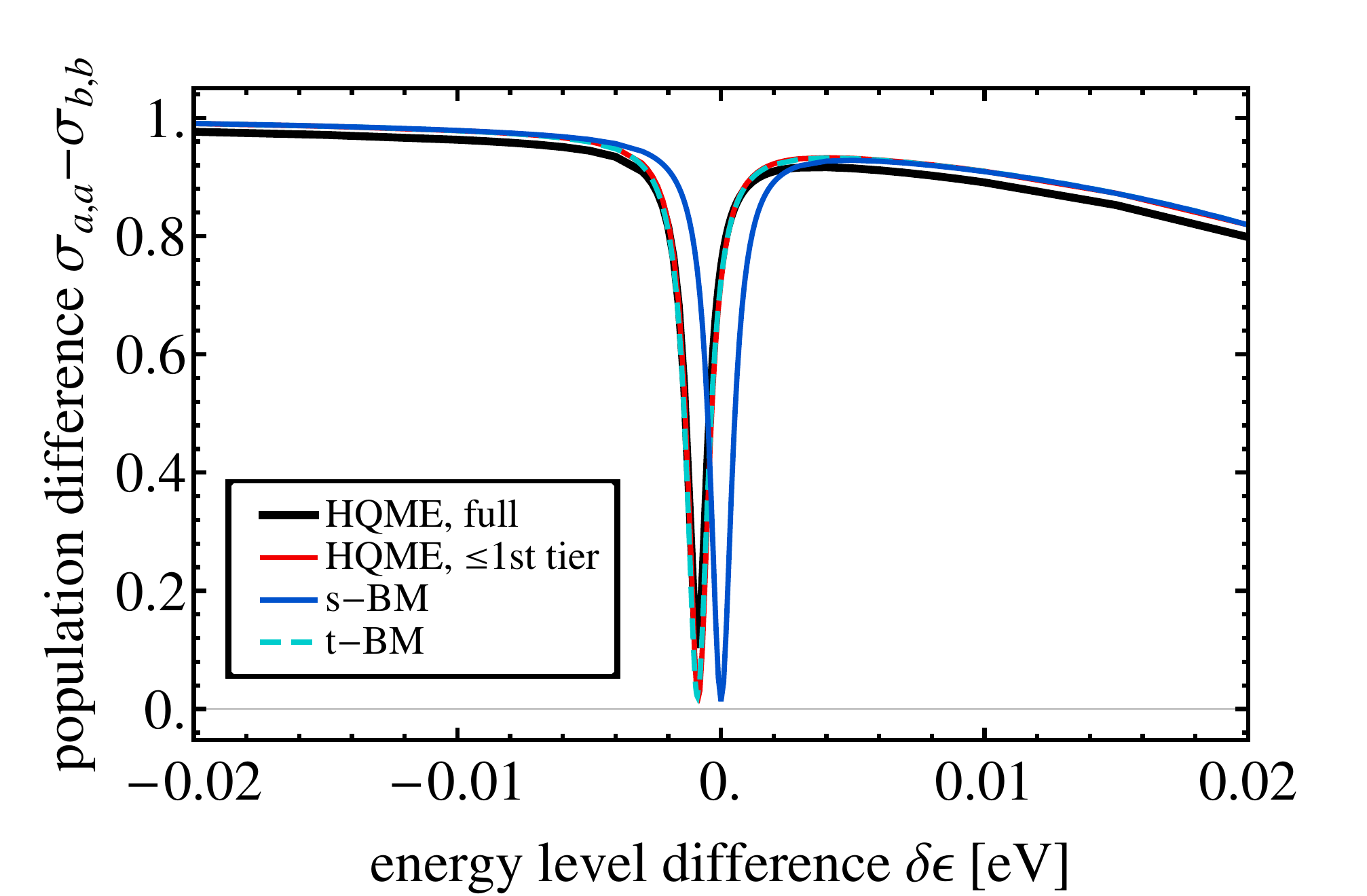}
}\\
\hspace{-0.5cm}(b) \\
\resizebox{\newwidth}{\newheight}{
\includegraphics{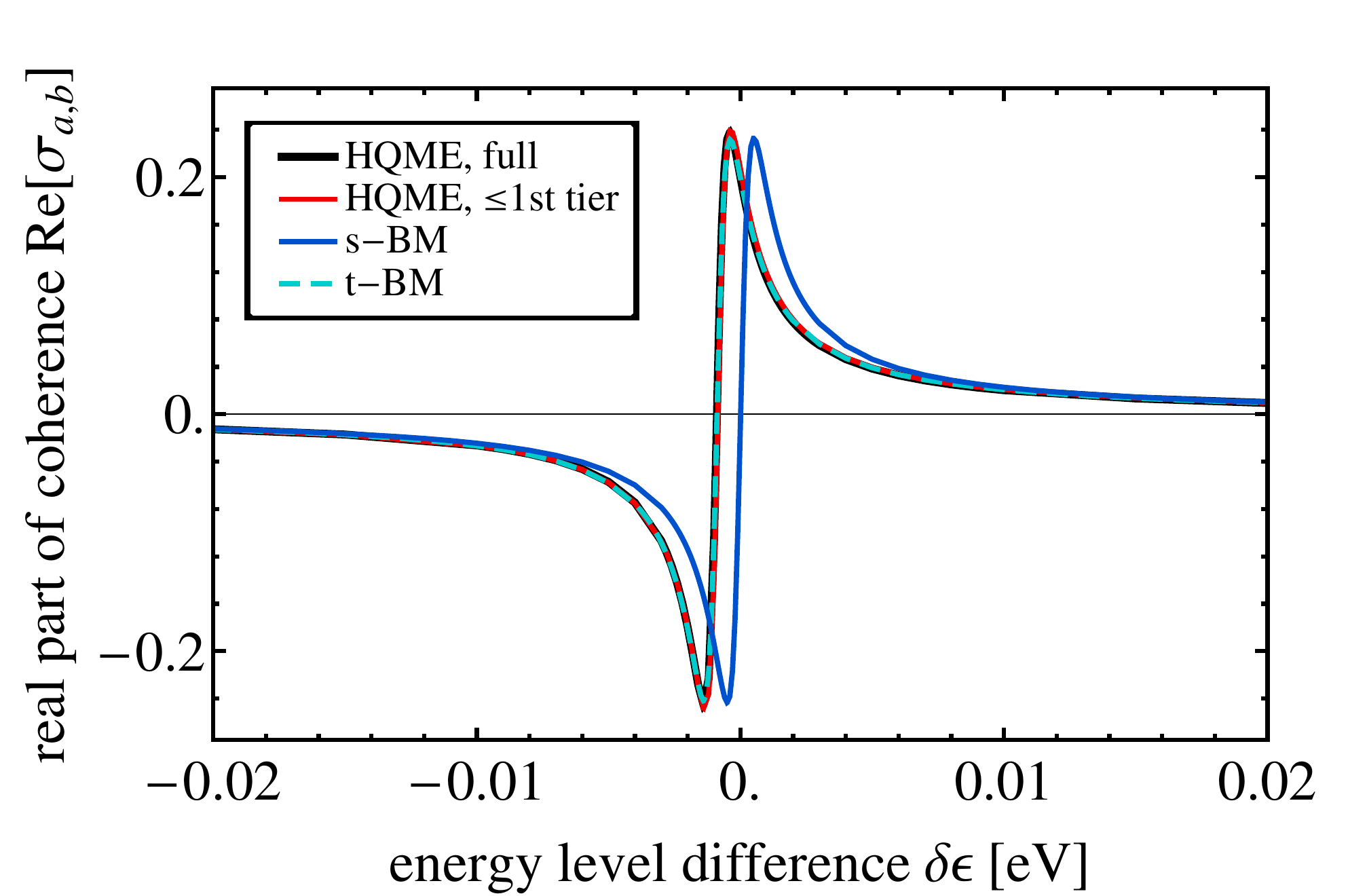}
}\\
\end{tabular}
\caption{(Color online) \label{deltaEdiff} 
Difference in the population of the dots $a$ and $b$ and the corresponding real part of the 
coherence $\sigma_{a,b}$ as a function of the energy level difference $\delta\epsilon$ in junction 
SERIAL at bias voltage $\Phi=0.1$\,V. The width of the dip structure is determined by the inter-dot 
coupling strength $\alpha=0.5$\,meV. 
}
\end{figure}

As we already pointed out, the origin of this behavior is the interaction-induced 
renormalizations $\Delta\epsilon_{a/b,\text{L/R}}$. To demonstrate this proposition, we vary the dot levels 
such that the effect of the $\Delta\epsilon_{a/b,\text{L/R}}$ is eventually cancelled. This is shown in 
Fig.\ \ref{deltaEdiff}, where the steady state population difference is depicted as a function of the energy 
level difference $\delta\epsilon$ (which is subtracted from $\epsilon_{a}$ and added to $\epsilon_{b}$). 
We see that the population difference becomes indeed minimal at values of $\delta\epsilon$ that correspond to 
a cancellation of the interaction-induced renormalizations $\Delta\epsilon_{a/b,\text{L/R}}$.

At this point, we like to highlight the non-trivial dynamics of this renormalization effect. To this end, 
we recall that the s-BM and t-BM scheme differ by principal value terms. For our systems of interest, these 
terms enter only the equation of motion of the coherence $\sigma_{a,b}$. The population 
difference does not occur, if the coherence, in particular the real part of the coherence, is neglected. 
This shows that the principal value terms encode not only static effects like a renormalization of energy 
levels but also relaxation mechanisms that are mediated by the coherence. In addition, we conclude that the 
effect is stable with respect to temperature as long as its contribution to the real part of the 
coherence $\sigma_{a,b}$ (cf.\ Sec.\ \ref{nonmarkovian}) is smaller than the one due to the 
interaction-induced renormalizations $\Delta\epsilon_{a/b,\text{L/R}}$. This is certainly the case if the 
energy separation of the eigenstates is much smaller than the thermal broadening.

These findings may also be interesting for quantum information processing 
\cite{Loss1998,Wiel2002,Sothmann2010,Baumgaertel2011,Zwanenburg2013}, 
as the coherence $\sigma_{a,b}$ between the dots can become sizeable ($\approx0.2$). Moreover, its value and sign 
can be controlled by the applied bias voltage. This is elucidated in more detail by Figs.\ \ref{biasdiff}(b), 
\ref{diffpop}(b) and \ref{deltaEdiff}(b), where the real part of the coherence is shown as a function of the applied voltage, 
energy level position $\epsilon_{0}$ and energy level difference $\delta\epsilon$, respectively. 
Once the bias voltage exceeds the thermal broadening, the real part of the coherence acquires its maximal value
before it decreases again when the system approaches the resonant transport regime. Its sign 
may be flipped by tuning the energy levels across the point where the population difference becomes minimal 
(and, finally, reaching the same population difference again). 
It is interesting to note at this point that the imaginary part of the coherence is given by the 
current, $\text{Im}\left[\sigma_{a,b}\right]\sim I$ (which we analyzed in detail in Ref.\ \onlinecite{Hartle2013b}). 
Thus, in junction SERIAL, the real and the imaginary part of the coherence may be disentangled.

\begin{figure}
\begin{center}
\includegraphics[width=\textwidth]{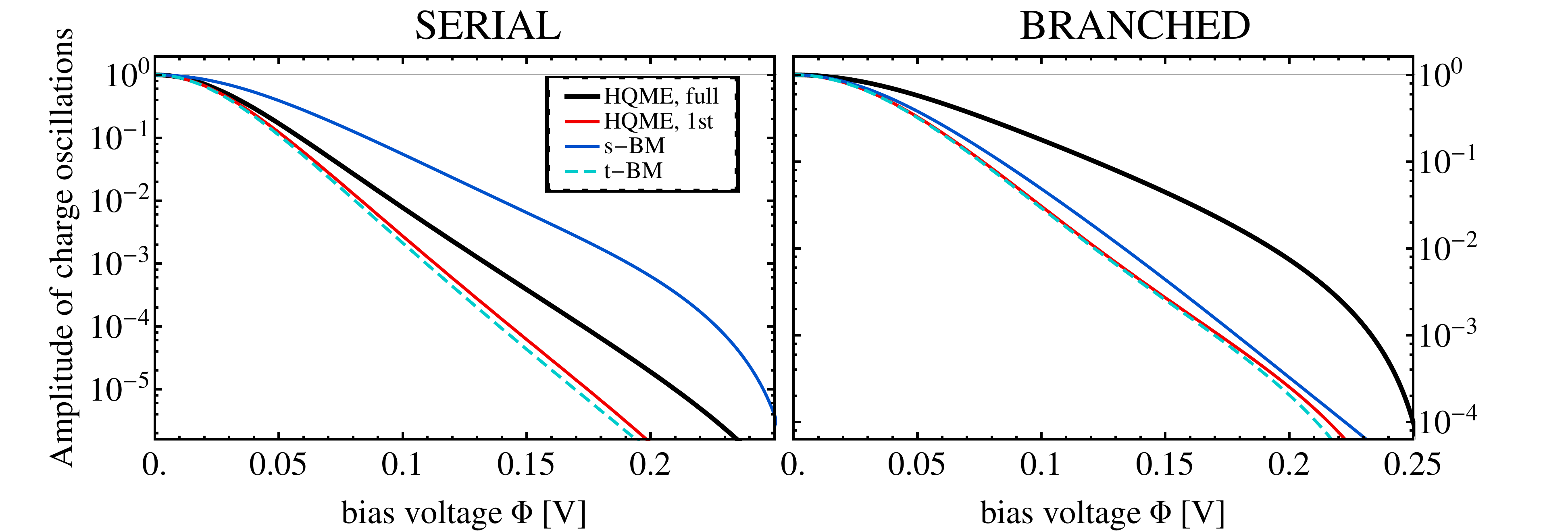}
\end{center}
\caption{(Color online) \label{amplitude} 
Normalized amplitude of the coherent charge oscillations in junction SERIAL (left plot) and junction BRANCHED (right plot) 
as a function of the applied bias voltage, starting from the asymmetric initial state $\sigma_{a,a}(t=0)=1$. 
To this end, a Fourier analysis of $\sigma_{a,a}(t)$ has been employed. 
}
\end{figure}

In contrast to junction SERIAL, system BRANCHED is much less affected by an external bias voltage. 
As can be seen in the right columns of Figs.\ \ref{NonequilibriumFig} and \ref{NonequilibriumFig1010}, 
the charge transfer oscillations between dots $a$ and $b$ decay on slightly shorter time scales and 
the corresponding amplitude becomes smaller. These findings can be understood as an increase of the 
effective temperature of the device. This picture is corroborated by the data shown in Fig.\ \ref{decaytimes}, 
which shows the decay times of the coherent charge oscillations in junction SERIAL and BRANCHED as a function 
of the applied bias voltage, and Fig.\ \ref{amplitude}, where the corresponding amplitudes are shown 
(starting from an initially asymmetric charge distribution). The data shows a clear exponential decrease 
of the decay times and the oscillation amplitude with an increasing bias voltage. Thereby, higher order 
processes seem to stabilize the coherent charge oscillations but, in fact, only increase the level broadening, 
that is the baseline of the dots effective temperature.

The exponential scaling of the amplitudes can be understood in more detail. To this end, we recall that the 
coherent charge oscillations require a different population of the two quantum dots. Such a population difference 
can emerge due to an initial asymmetry in the dots population or due to the geometry of the device 
(as, e.g., in junction BRANCHED). Thus, the difference in the dots population has to be present on time scales 
comparable to the period of the coherent charge oscillations. Initially, however, the population of the dots is 
governed by fast resonant tunneling processes between the electrodes and the dots. 
For junction SERIAL and the asymmetric initial condition $\sigma_{a,a}(0)=1$, 
the dominant decay channel is via hopping processes from the right lead onto dot b. The 
corresponding rate involves the Fermi funtion $f_{R}(\epsilon_{0}+U)\approx 
\text{exp}(-(\epsilon_{0}+U)/(k_{\text{B}}T)) \text{exp}(-\Phi/(2k_{\text{B}}T))$. 
For junction BRANCHED (and the asymmetric initial condition $\sigma_{a,a}(0)=1$), 
the dominant decay channel is via hopping processes from dot $a$ to the right lead, 
which occurs with a probability $\sim\text{exp}(-\epsilon_{0}/(k_{\text{B}}T)) \text{exp}(-\Phi/(2k_{\text{B}}T))$. 
The decay of the (normalized) amplitude can thus be estimated by 
$\text{exp}(-\Phi/(2k_{\text{B}}T))$, 
if only thermal broadening is taken into account 
(cf.\ the red, blue and turquoise lines in Fig.\ \ref{amplitude}), or by 
$\text{exp}(-\Phi/(2k_{\text{B}}T+\Gamma_{\text{L}}+\Gamma_{\text{R}}))$, if higher order processes are accounted for. 
This reasoning captures the scaling behavior that we observe in junction BRANCHED almost quantitatively. 
In junction SERIAL, interaction-induced renormalization effects lead to a slightly more complex behavior. 
This is evident from the different scaling behavior that is obtained from the s-BM scheme 
(see the left plot of Fig.\ \ref{amplitude}). Qualitatively, however, the behavior is very similar to the one of junction BRANCHED.

\section{Conclusion}
\label{conclusion}

Our results demonstrate the influence of an interaction-induced renormalization of energy levels 
on the coherent dynamics of a double quantum dot structure. This includes the formation of the 
steady-state coherence and populations and, on intermediate time scales, the period of coherent 
charge oscillations between the dots. In particular, the two quantum dots exhibit a pronounced 
population difference, which may be accessed in experiment non-invasively (\emph{e.g.}\ via point contacts), 
and a sizeable coherence, which is maximal in the non-resonant transport regime (cf.\ Fig.\ \ref{biasdiff}(b)).

To demonstrate these effects, we have focused on the regime where the structure holds a single 
electron on average. As a result, the build-up of the steady state is rather slow, allowing for 
long-lived intermediate dynamics which is governed by coherent processes. In this regime, 
transport processes strongly influence the charge distribution but coherent charge oscillations 
try to level off any asymmetry in the charge distribution. Due to this competition between transport 
and coherent dynamics, the population of the dots is very susceptible to small changes of the energy 
levels, in particular to interaction-induced renormalization effects. 
Thus, a way to detect interaction-induced renormalization and the corresponding coherent dynamics is 
to exploit its bias dependence. If, for example, the energy levels of a serial quantum dot system 
are aligned at zero bias, a pronounced population difference emerges at non-zero bias voltages, even 
though resonant transport is still suppressed 
(\emph{i.e.}\ $\Phi<2\text{Min}[\epsilon_{a/b},\epsilon_{a/b}+U]$, cf.\ Fig.\ \ref{biasdiff}(a)). 
In the same range of bias voltages, the coherence between the quantum dots is most pronounced and 
stabilized by the current that is flowing through the quantum dots. Its sign may be flipped by tuning 
the energy levels over a minimal population difference between the dots (cf.\ Fig.\ \ref{deltaEdiff}(b)).

Our analysis is based on numerically exact results, which are obtained by the hierarchical master equation technique 
\cite{Tanimura2006,Welack2006,Jin2008,Popescu2013,Hartle2013b}, and approximate results, which are based on both 
Born-Markov theory \cite{May02,Mitra04,Lehmann04,Harbola2006,Volkovich2008,Hartle2010}. The comparison of these 
results allowed us to reveal the physical mechanisms at work. They also demonstrate the need for numerically exact 
results, because the approximate results are spoiled by small (nevertheless unphysical) negative populations 
(cf., for example, the top left panel of Figs.\ \ref{NonMarkovianFig} and Fig.\ \ref{NonequilibriumFig}) and 
rather large errors in predicting the relevant time scales (see Figs.\ \ref{decaytimes} and \ref{timescales}). 
Moreover, we demonstrated that the hierarchical master equation technique is capable of describing the time 
evolution of an interacting quantum system on very long time scales. This includes both the times to reach the 
steady state ($\sim 10/\Gamma$--$10^{3}/\Gamma$, cf.\ Fig.\ \ref{timescales}) or the decay times of the coherent 
charge oscillations ($\sim 10/\Gamma$, cf.\ Fig.\ \ref{decaytimes}). This characteristics of the method is closely 
related to its time-local formulation (cf.\ Eq.\ (\ref{hierarcheom})).

\section*{Acknowledgements} 

We thank G.\ Cohen, J.\ Okamoto and C.\ Schinabeck for helpful comments. 
AJM is supported by the Basic Energy Sciences Division of the US Department of Energy under 
Grant No.\ DOE-FG02-04-ER046169. RH gratefully acknowledges financial support of the 
Alexander von Humboldt foundation via a Feodor Lynen research fellowship.

\appendix

\section*{Appendix: Parametrization of the correlation functions $C_{K,mn}^{s}$}
\label{appA}

To represent the correlation functions   
\begin{eqnarray} 
 C^{s}_{K,mn}(t) &=& \int_{-\infty}^{\infty}\frac{\text{d}\omega}{2\pi}\, \text{e}^{si\omega t} 
 \Gamma_{K,mn}^{s}(\omega) f^{s}_{K}(\omega), 
\end{eqnarray}
by a set of exponentials, we first express the distribution functions $f^{s}_{K}(\omega)$ by a sum over poles 
\begin{eqnarray}
 f^{s}_{K}(\omega) &=& \frac{1}{2} - s \frac{1}{4} \sum_{p} \frac{ R_{p} }{x+i E_{p}}.
\end{eqnarray}
To this end, we employ the Pade approximation \cite{Hu2010,Hu2011}. 
Thus, according to Ref.\ \onlinecite{Ozaki2007}, 
the pole positions $E_{p}$ are identical with 
the eigenvalues of a tridiagonal matrix with the coefficients 
\begin{eqnarray}
 A_{ij} = \delta_{i,j+1} \frac{1}{2\sqrt{ (2i+1) (2i-1) }} + \delta_{i,j-1} \frac{1}{2\sqrt{ (2j+1) (2j-1) }}. 
\end{eqnarray}
The weights $R_{p}$ are given by 
\begin{eqnarray}
 R_{p}  &=& E_{p}^{2} \left\vert \langle p \vert 1 \rangle \right\vert^{2} ,
\end{eqnarray}
where $\langle p \vert 1 \rangle$ denotes the overlap of the 
$p$th eigenvector $\vert p \rangle$ with the vector 
$\vert 1 \rangle = (1,0,0,0,...)^{\text{T}}$. 
The next step is to represent 
the level-width functions $\Gamma_{K,mn}^{s}(\omega)$ by a similar expression. 
This can be done, for example, using a Meir-Tannor parametrization scheme \cite{Tannor1999,Welack2006,Jin2008}, 
but is obsolete for the Lorentzian conduction bands that we employ in this work (see Eq.\ (\ref{lorbands})). 
Finally, the amplitudes $\eta_{K,mn,p}^{s}$ and frequencies $\omega_{K,p}^{s}$ are obtained straightforwardly  
via contour integration.

Throughout this work, we have used $100$ Pade poles in order to get converged results.  
Thereby, we reduce the number of auxiliary 
operators $\sigma^{(\kappa)}_{j_{1}..j_{\kappa}}(t)$ to a practical level 
using the systematic truncation scheme that we developed in 
Ref.\ \onlinecite{Hartle2013b}. Thus, the actual number of Pade poles is less decisive for 
the numerical effort, as we briefly exemplify in Tab.\ \ref{tabnum}. Note that 
it is beneficial to use a low number of poles, because the frequencies $\omega_{K,p}^{s}$ increase 
with the pole index $p$ requiring a higher resolution of the time axis.

\begin{table}
\begin{center}
\begin{tabular}{|c|*{7}{ccc|}}
\hline 
\# of Pade poles: && 40 &&& 60 &&& 100 &&& 200 &&& 400 &&& 800 &\\ \hline 
\# of ADOs: && 7653  &&& 11019  &&& 12863  &&& 14551 &&& 15711&&& 15822 &\\
max.\ tier level: && 4  &&& 4  &&& 4  &&& 4 &&& 4 &&& 4 &\\
\hline 
\end{tabular}
\end{center}
\caption{\label{tabnum} Number of auxiliary operators for an increasing 
number of Pade poles that are included in our calculations. Due to our specific 
truncation scheme (see appendix of Ref.\ \onlinecite{Hartle2013b}), 
which allows a systematic reduction of the number of auxiliary operators 
$\sigma^{(\kappa)}_{j_{1}..j_{\kappa}}(t)$, the numerical effort levels off 
with an increasing number of Pade poles. 
}
\end{table}

\end{document}